\documentclass[proof]{pasj01}
\draft
\begin{document} 
\Received{}
\Accepted{}

\title{Investigation of the upper atmosphere in ultra-hot Jupiter WASP-76 b with high-resolution spectroscopy}

\author{Kiyoe Kawauchi$^{1,2}$, Norio Narita$^{3,4,5,2}$, Bun'ei Sato$^6$, Yui Kawashima$^{7,8}$}

\altaffiltext{1}{Department of Multi-Disciplinary Sciences, Graduate School of Arts and Sciences, The University of Tokyo, 3-8-1 Komaba, Meguro, Tokyo 153-8902, Japan}
\altaffiltext{2}{Instituto de Astrof\'\i sica de Canarias (IAC), 38205 La Laguna, Tenerife, Spain}
\altaffiltext{3}{Komaba Institute for Science, The University of Tokyo, 3-8-1 Komaba, Meguro, Tokyo 153-8902, Japan}
\altaffiltext{4}{JST, PRESTO, 3-8-1 Komaba, Meguro, Tokyo 153-8902, Japan}
\altaffiltext{5}{Astrobiology Center, 2-21-1 Osawa, Mitaka, Tokyo 181-8588, Japan}
\altaffiltext{6}{Department of Earth and Planetary Sciences, Tokyo Institute of Technology, 2-12-1 Ookayama, Meguro-ku, Tokyo 152-8551, Japan} 
\altaffiltext{7}{Cluster for Pioneering Research, RIKEN, 2-1 Hirosawa, Wako, Saitama 351-0198, Japan}
\altaffiltext{8}{SRON Netherlands Institute for Space Research, Sorbonnelaan 2, 3584 CA Utrecht, The Netherlands}


\KeyWords{planets and satellites: atmospheres --- planets and satellites: individual (WASP-76 b) --- planets and satellites: composition} 

\maketitle

\begin{abstract}
Alkali metal lines are one of the most important key opacity sources for understanding exoplanetary atmospheres because the Na I resonance doublets are thought to be the cause of low albedo, as the alkali metal's wide line wings absorb almost all of the incoming stellar irradiation.
High-resolution transmission spectroscopy of Na absorption lines can be used to investigate the temperature of the thermosphere of hot Jupiters, which is increased by stellar X-ray and EUV irradiation.
We applied high-resolution transmission spectroscopy to the ultra-hot Jupiter WASP-76 b with the High Dispersion Spectrograph (HDS) on the Subaru 8.2~m telescope. 
We report the detection of strong Na D excess absorption with line contrasts of 0.42 $\pm$ 0.03 \% (D1 at 5895.92 \mbox{\AA}) and 0.38 $\pm$ 0.04 \% (D2 at 5889.95 \mbox{\AA}), FWHMs of 1.63 $\pm$ 0.13 \mbox{\AA} (D1) and 1.87 $\pm$ 0.22 \mbox{\AA} (D2), and EWs of (7.29 $\pm$ 1.43) $\times$ 10$^{-3}$ \mbox{\AA} (D1) and (7.56 $\pm$ 2.38) $\times$ 10$^{-3}$ \mbox{\AA} (D2). These results show that the Na D absorption lines are shallower and broader than those in previous work, whereas the absorption signals over the same passband are consistent with those in previous work.
We derive the best-fitted isothermal temperature of 3700 K (without rotation) and 4200 K (with rotation).
These results suggest the possibility of the existence of a thermosphere because the derived atmospheric temperature is higher than the equilibrium temperature ($\sim$ 2160 K).

\end{abstract}

\section{Introduction}
Previous observations have found a number of hot Jupiters, which are gaseous planets that orbit close to a star, unlike our solar system. Although hot Jupiters can be relatively easily detected and investigated because of their large signal, their formation and evolution are not fully understood.
It is thus important to obtain detailed information about their atmospheric composition and temperature structure.
 
In the last two decades, our understanding of exoplanetary atmospheres has advanced considerably through spectroscopic observations during planetary transit events. Such observations are referred to as transmission spectroscopy. Alkali metal lines are one of the most important key opacity sources because the Na I resonance doublets are thought to be the cause of generally low albedos, as the alkali metal's wide line wings absorb almost all of the incoming stellar irradiation \citep{Seager&Sasselov2000, Brown+2001, Barman+2001,Hubbard+2001}.

High-resolution transmission spectroscopy can be used to investigate the upper atmosphere based on the centers of absorption lines. In particular, Na absorption in the line core makes it possible to investigate a thermosphere where the temperature is increased by stellar X-ray and extreme ultraviolet (EUV) irradiation and there is atmospheric escape \citep{Lammer+2003,Yelle2004,Vidal-Madjar+2011,Koskinen+2013}.
Na absorption has been detected with high-resolution spectroscopy for ten hot Jupiters and one hot Neptune (e.g., \cite{Wyttenbach+2015,Wood+2011,Casasayas-Barris+2017,Wyttenbach+2017,Casasayas-Barris+2018,Burton+2015,Seidel+2019,Hoeijmakers+2019,Chen+2020,Hoeijmakers+2020,Seidel+2020b}); however, there is some differences in these results that are not fully understood because the parameters vary between planets'.
To better understand the atmospheric structures of hot Jupiters and explain these differences, it is necessary to apply high-resolution spectroscopy to observe the Na absorption lines for a greater number of hot Jupiters.

There is a classification system that roughly divides hot Jupiters into two types based on the incident flux (equilibrium temperature of a planet) \citep{Fortney+2008}.
For an ``ultra-hot Jupiter" whose equilibrium temperature is over 2000 K, it is predicted that TiO/VO absorption is governed by the optical wavelength and that Na absorption lines cannot be detected \citep{Fortney+2010}. 
However, recent observations using transit photometry and low/high resolution spectroscopy detected many alkali lines and/or the Balmer series of H and/or TiO/VO in some ultra-hot Jupiters, revealing differences in their atmospheres (e.g., \cite{Sing+2013,Burton+2015,Evans+2016,Sedaghati+2017, Jensen+2018,Casasayas-Barris+2018,Wyttenbach+2020,Hoeijmakers+2020}). 
To investigate these differences, observations of the atmosphere in more ultra-hot Jupiters are needed.

WASP-76 b, which is around an F-type star, was confirmed by \citet{West+2016}.
This planet is an inflated ultra-hot Jupiter with a radius of 1.854 $\rm{R_J}$, a mass of 0.894 $\rm{M_J}$, an equilibrium temperature of 2228 K, and an orbital period of  $\sim$ 1.8 days. The parameters of the WASP-76 system are listed in table~\ref{table:param}. Because the WASP-76b system has a bright host star and a large scale height, its atmosphere can be investigated. 

In 2019, two studies reported the detection of Na absorption lines in the atmosphere of WASP-76 b with high-resolution spectroscopy \citep{Seidel+2019,Zak+2019}. These studies used the same data (observed with HARPS in 3 nights) but different analysis methods.
Both detected Na absorption, but the depth and FWHM values were different.
Furthermore, \citet{Ehrenreich+2020} confirmed the existence of iron, which condenses at the nightside, from the asymmetry of the absorption signal of atomic iron obtained with the Echelle Spectrograph for Rocky Exoplanets and Stable Spectroscopic Observations (ESPRESSO) at the European Southern Observatory's Very Large Telescope (VLT).
In space, the low-resolution spectra (transmission and emission spectra) of WASP-76 b were observed several times with STIS, Wide Field Camera 3 (WFC3) of the Hubble Space Telescope (HST), and Spitzer. The results were used to calculate a temperature-pressure profile at the terminator and dayside, and the existence of TiO and H$_2$O were detected in the transmission spectra, a strong CO emission feature was found in the secondary eclipse spectra, and sodium was found at 2.9$\sigma $ significance. \citep{Edwards+2020,Fu+2020arXiv,Lothringer+2020,vonEssen+2020}.
In addition, it is found that WASP-76 has a companion star whose temperature is 4824 K and which is separated by about 0".436 \citep{Bohn+2020,Southworth+2020}. Because this companion cannot separate from the host star in photometry, it affects HST data.
Most recently, the transmission spectra of two primary transits were observed with ESPRESSO; the atomic absorption lines of Li I, Na I, Mg I, Ca II, Mn I, K I, and Fe I were detected \citep{Tabernero+2021}.

Thanks to the previous studies, many atoms and molecules are detected in the atmosphere of WASP-76b. The existence of the temperature inversion is estimated from the detection of atoms and molecules such as TiO/VO, but it is not confirmed in the retrieval fitting with transmission spectra of HST. In high resolution, \cite{Seidel+2019} tried to fit their data with the theoretical transmission spectrum, but they could not adjust the line profile.

In the present study, we show the high-resolution transmission spectra of WASP-76 b observed with the Subaru 8.2 m telescope and High Dispersion Spectrograph (HDS; \cite{Noguchi+2002}) and fit them with our theoretical transmission spectra considering equilibrium chemistry, namely thermal ionization of sodium. The rest of this paper is organized as follows. In Sections 2 and 3, we describe our observations and the procedure used for data reduction, respectively. In Section 4, we show the results of the transmission spectrum and atmospheric absorption. In Section 5, we discuss the effect of instrumental variation and conduct a comparison with model spectra. In Section 6, we summarize this work.

\begin{table}[htbp]
\begin{center}
\tbl{Parameters of WASP-76 system}{
\begin{tabular}{lc}
\hline \hline
 Parameter & Value \\
\hline
 Star & WASP-76 \\
 V mag & 9.5\\
 Spectral type & F7$^{*}$\\
 $R_{\rm s} (\rm{R_{sun}})$ & 1.756 $\pm$ 0.071$^{\dagger}$\\
 $T_{\rm eff}$ (K) & 6329 $\pm$ 65$^{\dagger}$\\
 \hline
 Transit &  \\
 $T_{\rm 0}$ (UTC) & 2456107.85507 $\pm$ 0.00034$^{*}$ (HJD)\\
 $P$ (day) & 1.809886 $\pm$ 0.000001$^{*}$\\
Duration (h) & 3.694 $\pm$ 0.019$^{*}$\\
Impact parameter : $b$ & 0.14 $^{+0.11}_{-0.09}$$^{*}$\\
Semimajor axis : $a$ (AU) & 0.0330 $\pm$ 0.0005$^{*}$\\
\hline
Planet & WASP-76 b\\
$M_{\rm P} (\rm{M_{J}})$ & 0.894 $^{+0.014}_{-0.013}$$^{\dagger}$\\
$R_{\rm P} (\rm{R_{J}})$ & 1.854 $^{+0.077}_{-0.076}$$^{\dagger}$\\
$T_{\rm P,A=0}$ (K) & 2228 $\pm$ 122$^{\dagger}$\\
$g$ = G$M_{P}/R_{P}^2$ (cm $\rm{s^{-2}}$) & $\sim $ 645 \\
$H $ (km) & $\sim$ 1249\\
\hline
\end{tabular}}\label{table:param}
\end{center}
\begin{flushleft}
Notes. $^{(*)}$ \citet{West+2016}. \ $^{(\dagger)}$ \citet{Ehrenreich+2020}. \ $H$ is calculated  assuming a mean molecular weight $\mu$ = 2.3.
\end{flushleft}
\end{table}

\section{Observations}
On UT September 10, 2018, we observed the F-type star WASP-76 during the
transit of its hot Jupiter WASP-76 b with Subaru/HDS. The wavelength range
was set to cover 5430--8100 \mbox{\AA}, which is a nonstandard setup for HDS, using a red cross-disperser grating with a central wavelength of 6700 \mbox{\AA} to observe the Na (D1 = 5895.92 \mbox{\AA}; D2 = 5889.95 \mbox{\AA}) and K absorption lines simultaneously. The slit width was set to 0.4", corresponding to a spectral resolution of $\sim$ 90,000.

Because the transit duration of the planet is about 3.7 h from 10:41 UT to 14:23 UT, observation started at 8:22 UT and finished at 15:41 UT. We obtained a total of 35 out-of-transit spectra and 35 in-transit spectra of WASP-76 b with an exposure time of 300 s. The signal-to-noise ratio (SNR) of the extracted spectra was about 86 $\rm pix^{-1}$ on average and about 23 $\rm pix^{-1}$ for the line core of the Na D lines at each frame.

To obtain the telluric spectra, we observed the rapid rotator HR 8634 (B8V, $v \sin i = $185 km s$^{\rm -1}$; \cite{Glebocki+2000}) from 5:19 UT to 8:11 UT.  We obtained 180 exposures of 5 s each for the star. The SNR of the extracted spectra was about 158 $\rm pix^{-1}$ on average at each frame.

\begin {figure} [htbp]
\begin{center}
 \includegraphics[width = 8.0cm] {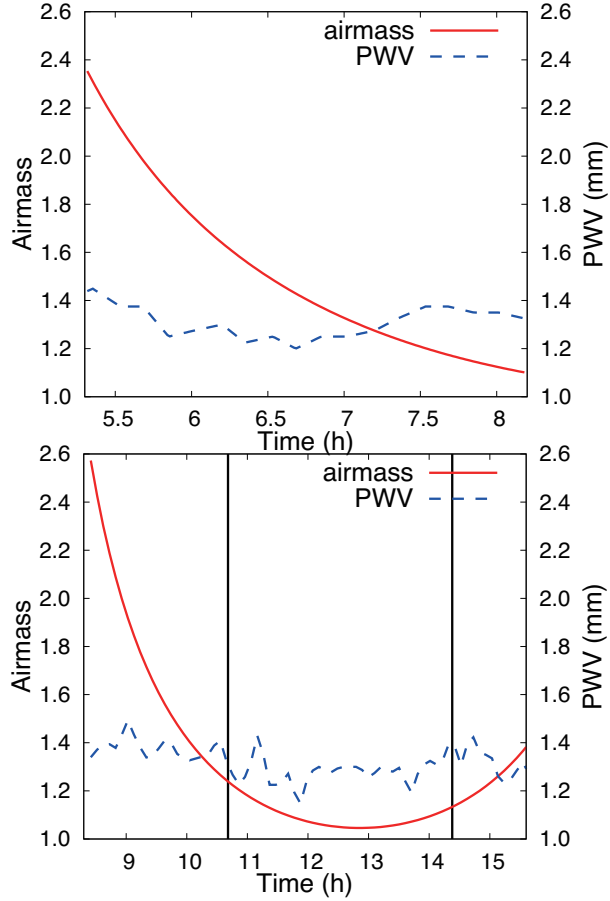}
\end{center}
\caption{Earth's atmospheric condition during the observation of HR 8634 (top) and WASP-76 (bottom) on UT September 10, 2018. The vertical lines represent the start (left) and end (right) of transit. The variations in the airmass and PWV are represented by the red and blue lines, respectively.}
\label{fig:weatherWASP76}
\end{figure}

The Earth's atmospheric conditions during the observation of WASP-76 and HR 8634 are shown in figure~\ref{fig:weatherWASP76}. 
The precipitable water vapor (PWV), which refer to the depth of water vapor in a column of the atmosphere, was measured by the Caltech Submillimetre Observatory radiometer at 225 GHz from the linear relation between the measurable PWV in millimeters and the opacity at 225 GHz \citep{Dempsey+2013}. 
The average PWV was about 1.32 mm, which means that the atmosphere is dry. This value was relatively stable during the observations. 
Table~\ref{table:log} summarizes the observation log.

\begin{table}[htbp]
\tbl{Observation log}{
\begin{tabular}{cccccc}
\hline \hline
 Object & Date & Obtained spectra & Exp. time & Airmass & PWV \\
 & (UT) & in-transit + out-of-transit & (s) & & (mm) \\
\hline
WASP-76 & 2018/09/10 & 35 + 35 & 300 & 1.05--2.57 & 1.32 $\pm$ 0.07\\
\hline
\end{tabular}}\label{table:log}
\end{table}

\section{Data reduction}
To extract the transmission spectra, the one-dimensional spectral flux as a function of wavelength was extracted from the two-dimensional images recorded on a CCD detector and the telluric spectrum, stellar spectrum, and instrumental variation were removed.
The first data reduction was performed using the Image Reduction and Analysis Facility (IRAF)\footnote[1]{IRAF is distributed by the U.S. National Optical Astronomy Observatories, which are operated by the Association of Universities for Research in Astronomy, Inc., under a cooperative agreement with the National Science Foundation.}.
Other procedures are described below.

\subsection{Telluric correction}
Telluric absorption lines were superposed onto the stellar spectrum observed from the ground. Many telluric water vapor absorption lines appear around Na D lines. 
There are three main methods for removing these telluric absorption lines. One uses the spectra of a rapidly rotating star (a rapid rotator) (e.g., \cite{Winn+2004, Kawauchi+2018}), one uses the telluric lines in out-of-transit spectra (e.g., \cite{Astudillo-Defru&Rojo2013, Wyttenbach+2015}), and one uses "molecfit" program \citep{Smette+2015,Kausch+2015} which combines model spectra of Earth's atmosphere with the local weather profile (e.g., \cite{Hoeijmakers+2020,Seidel+2020b}). 

We used the first method in the present study. We observed the rapid rotator HR 8634, which is known to have relatively few interstellar absorption lines, for about 3 h on the same day as WASP-76 was observed and obtained about 180 spectra at various airmasses.
These spectra are suitable for obtaining the telluric absorption lines because the weather conditions were stable during the observation of WASP-76 (figure~\ref{fig:weatherWASP76}). Using the spectra of the rapid rotator, we measured the optical depth of the telluric absorption lines during observations from the relationship between airmass ($AM$) and telluric absorption lines ($T_{AM}$) which is expressed as
\begin{equation}
\ln T_{AM}(\lambda) = \tau_{\rm \lambda}  AM
\end{equation}
where $\tau_{\rm \lambda}$ is the optical depth at a given wavelength.
We removed the telluric spectra from the spectra of WASP-76 using the obtained optical depth.

\begin{figure}[htbp]		
 \begin{center}
 \includegraphics[width = 16.0cm] {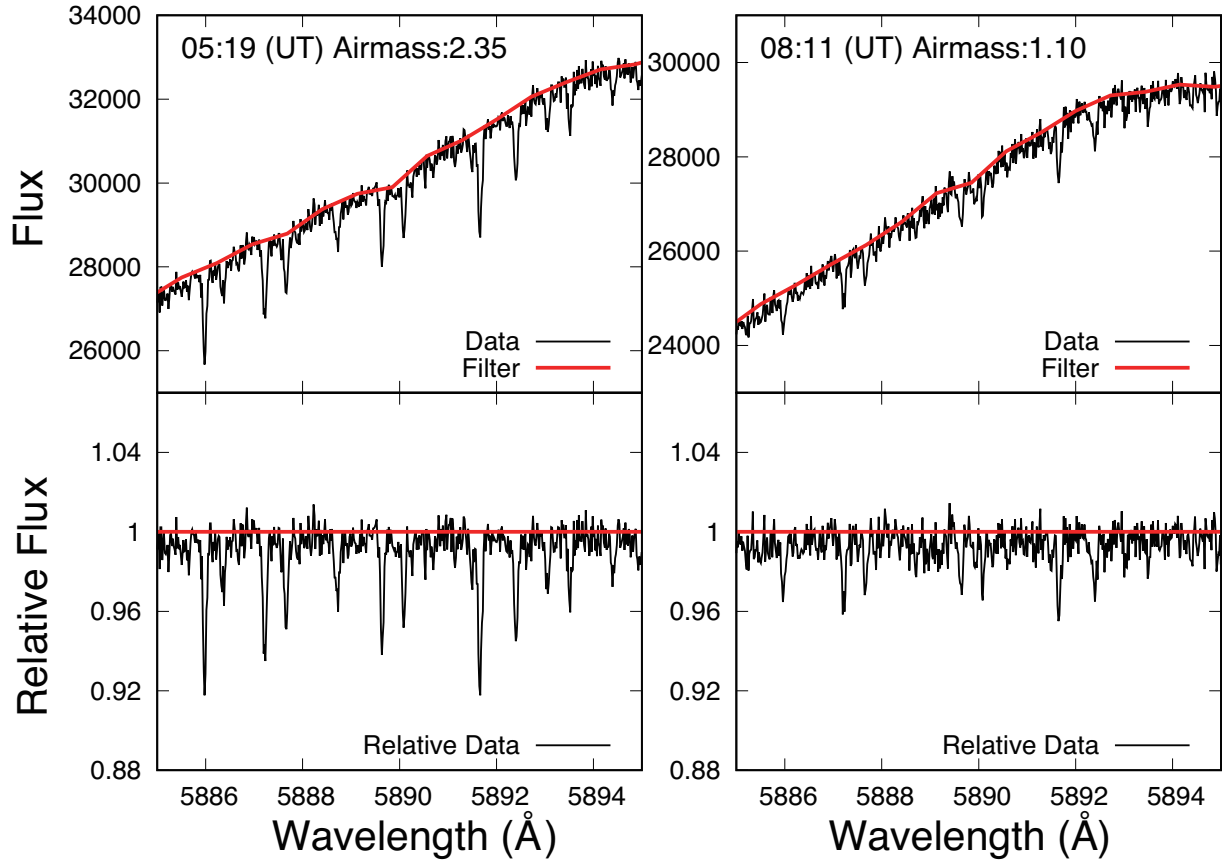}
\end{center}
\caption{{The top panels indicate the spectra of the reference star HR8634 (black) and the calculated filter to remove the continuum (red) at two different time. The bottom panels indicate the relative spectra (black) and the horizontal line of 1.}}
\label{fig:Blazetelluric}
\end{figure}

\begin{figure}[htbp]		
 \begin{center}
 \includegraphics[width = 16.0cm] {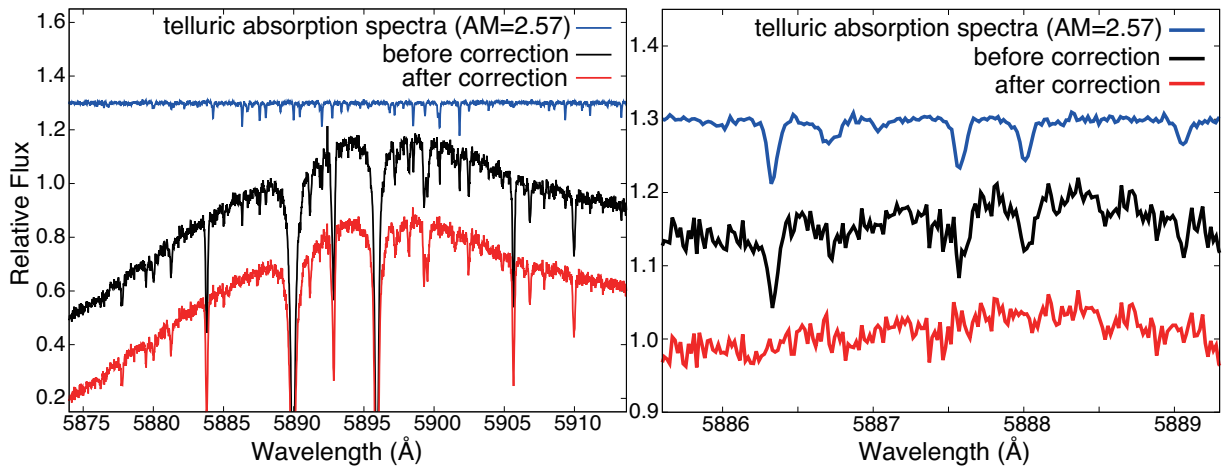}
\end{center}
\caption{{One of the HDS spectra of WASP-76b before (black) and after (red) telluric correction. The blue lines is the telluric absorption spectra. The right panel indicates the zoom of the left panel.  }}
\label{fig:telluricKAI}
\end{figure}

However, because the rapid rotator has a broad absorption feature, the continuum feature is not the same as those for neighboring orders. Therefore, we could not use the correction method used for the stellar spectra.
To create normalized spectra, {we divided the spectra every 40 pixels (5 $\times$ FWHM), selected the 9th largest value (80 \% from the bottom) as the value of the central wavelength, and fit these values using cubic spline (figure~\ref{fig:Blazetelluric}). The divided pixel and the used value are decided by changing at each FWHM ($\sim$ 0.15 \mbox{\AA}) and at each 10 \%.}
Based on these normalized spectra, we measured the optical depth around Na absorption lines. {Using this optical depth, we corrected the telluric spectra in WASP-76b spectra to the noise level (figure~\ref{fig:telluricKAI}).}

\subsection{Correction for instrumental variation}
The stellar spectrum observed with an echelle spectrograph such as Subaru/HDS has a blaze function for each observed order. However, in reality, this blaze function also has a time variation. This time variation can reduce the accuracy of flux correction using a standard star. It is considered that this time variation is caused by the changes in the optical path of the telescope and is related to the telescope position during observation and/or its focal length. However, because this time variation is not fully understood, it cannot be corrected for systematically. 

The target spectrum can be fitted using the continuum region to remove the effects of the blaze function. However, the blaze function cannot be fitted with high accuracy around broad and deep absorption lines such as Na D lines. For our observations, high-accuracy flux correction is necessary because the absorption lines of the exoplanet atmosphere are quite sensitive to differences in the continuum level. Therefore, we used the correction method adopted by \citet{Winn+2004} and \citet{Narita+2005b} because it has a relatively high accuracy {(the remaining instrumental effect is $\sim$ 0.1$\%$ with 2 \mbox{\AA} passband reported by \cite{Narita+2005b})}.

In their analysis, \citet{Winn+2004} used all 4100 pixels in the order while \citet{Narita+2005b} only used only about 1000 pixels around the Na D absorption lines to prevent the dilution of real signals.
In our analysis, we used about 2300 pixels to cover the region of the reference band {(5874.89-5886.89 \mbox{\AA} and 5898.89-5910.89 \mbox{\AA})}.

\begin{figure}[htbp]	
 \begin{center}
 \includegraphics[width = 8.0cm] {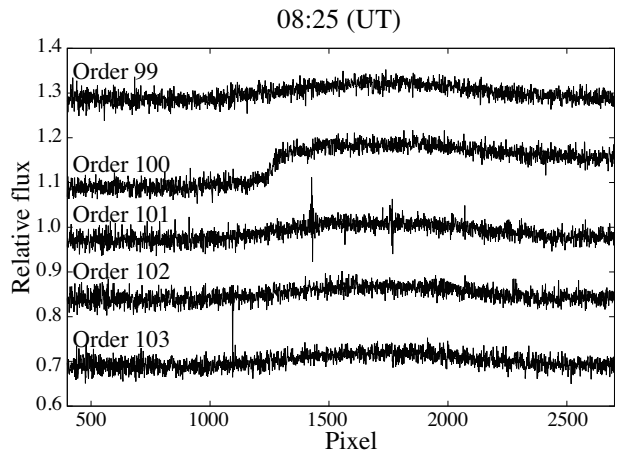}
\end{center}
\caption{Ratio between the template spectrum and the spectra for orders 99, 100, 101, 102, and 103. The variation of order 100 is different from those of the other orders.}
\label{fig:timevariationofblazefunc}
\end{figure}

We estimated the temporal variation by interpolating those obtained for neighboring orders based on the ratio between the spectrum that combines all spectra and the spectra at a given time.
To estimate the temporal variation with more accuracy than that in previous studies, we removed outliers from neighboring orders and smoothed these orders by an average of 40 pixels.
We used orders 99 and 103 to correct for the time variation of order 101, which has Na D absorption lines, because order 100 contains different time variations compared to the other orders (figure~\ref{fig:timevariationofblazefunc}).

\begin{figure}[htbp]	
 \begin{center}
   \includegraphics[width = 8.0cm] {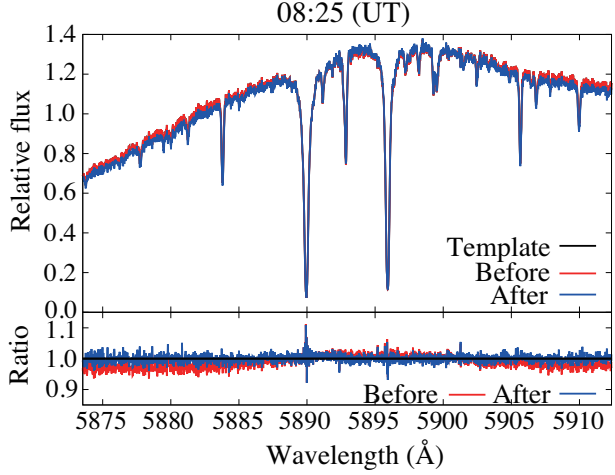}
  \end{center}
 \caption{Blaze correction at certain time points. The upper panel shows the template spectrum (black) and the spectra before (red) and after (blue) the blaze function correction. The bottom panel shows the ratio of the template spectrum to the spectra before (red) and after (blue) the blaze function correction. The black line in the bottom panel indicates a ratio of one.}
 \label{fig:bcWASP76}
\end{figure}

Figure~\ref{fig:bcWASP76} shows the original and corrected spectra and the ratio between the template spectrum and the spectrum at a given time.
These results show that all spectra are well corrected (the residual was around zero for all wavelengths).

\subsection{Removal of stellar spectrum}
The out-of-transit spectrum after telluric and instrumental variation corrections includes only the stellar spectrum.
We combined all corrected out-of-transit spectra to obtain the template stellar spectrum $F_{\rm star} (\lambda)$.
Under the assumption that the stellar spectrum did not vary during observation,
we removed the stellar spectrum by dividing each spectrum $F (\lambda,t)$ by $F_{\rm star} (\lambda)$:
\begin{equation}
\tilde{F} (\lambda, t) = F (\lambda, t) / F_{\rm star} (\lambda).
\end{equation}

\section{Results}
\subsection{Transmission spectrum}
The transmission signal was too faint to detect absorption lines. We thus combined the in-transit spectra to obtain the transmission spectrum. However, the absorption lines of the planet's atmosphere are shifted during transit because the planet orbit around its star.

The variation of the planet radial velocity ($\Delta V_{\rm R}$) can be expressed as
\begin{equation}
\Delta V_{\rm R} = \frac{2 \pi a}{P}  \times \sin \left[ 2 \pi \left( \frac{t-t_{\rm c}}{P} \right) \right]
\end{equation}
where $P$ is the orbital period, $a$ is the semi-major axis, $t$ is the time of observation of each frame and $t_{\rm c}$ is the time of observation of the mid-transit \citep{Khalafinejad+2017}.
The radial velocity of WASP-76 b changed from +50.8 kms$^{-1}$ to -50.3 kms$^{-1}$ during transit.
We shifted each spectrum by the value calculated using the above equation and combined the shifted transmission spectra during transit using the following equation:
\begin{equation}
T(\lambda) = \sum_{\rm in} \tilde{F}(\lambda, t_{\rm in})|_{\rm planet RV shift}
\end{equation}

\begin {figure*} [htbp]
\begin{center}
 \includegraphics[width = 16.0cm]{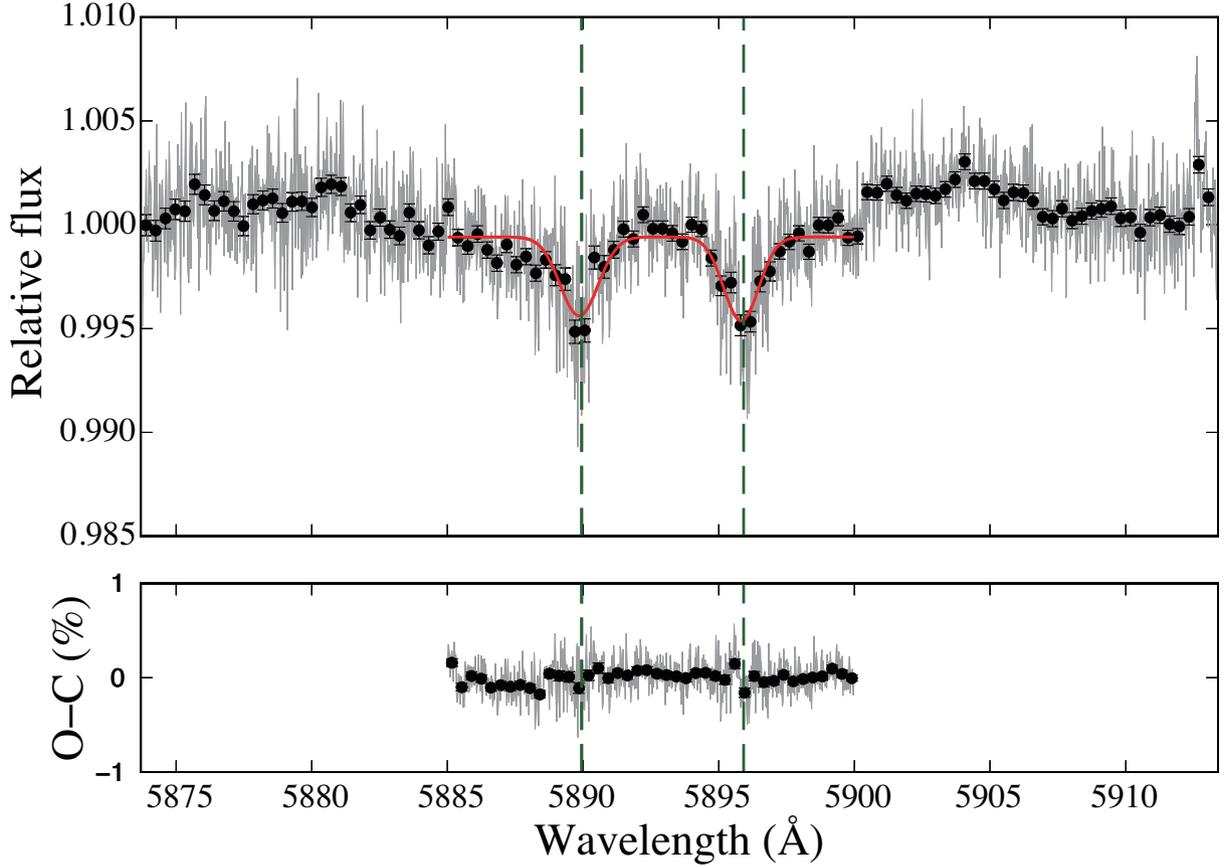}
\end{center}
\caption{Transmission spectrum corrected by planet radial velocity {before correction of a wide trend}. Red line indicates the fitting result obtained with a flat line and two Gaussians between 5885 and 5900 \mbox{\AA} (we could not fit between 5873.7 and 5913.4 \mbox{\AA}). The bottom panel shows the residuals of fitting.}
\label{fig:resultWASP76_prv}
\end{figure*}

 This allowed us to visually identify the Na D lines (figure~\ref{fig:resultWASP76_prv}). We estimated the transmission spectrum with a Gaussian fitting of each Na D line using Markov chain Monte Carlo (MCMC) analysis using the ensemble sampler emcee \citep{Foreman-Mackey+2013}. 
 We fit the observed spectrum with a flat line and two Gaussians between 5885 and 5900 \mbox{\AA} and measured line contrasts of 0.40 $\pm$ 0.03 \% (D1) and 0.38 $\pm$ 0.05 \% (D2) and full width at half maximum (FWHM) values of 1.52 $\pm$ 0.13 \mbox{\AA} (D1) and 1.65 $\pm$ 0.29 \mbox{\AA} (D2). 
 We also calculated equivalent widths (EWs) of (6.47 $\pm$ 1.42) $\times$ 10$^{-3}$ \mbox{\AA} (D1) and (6.66 $\pm$ 3.13) $\times$ 10$^{-3}$ \mbox{\AA} (D2).

\begin{figure*}[htbp]	
\begin{center}
 \includegraphics[width = 16.0cm]{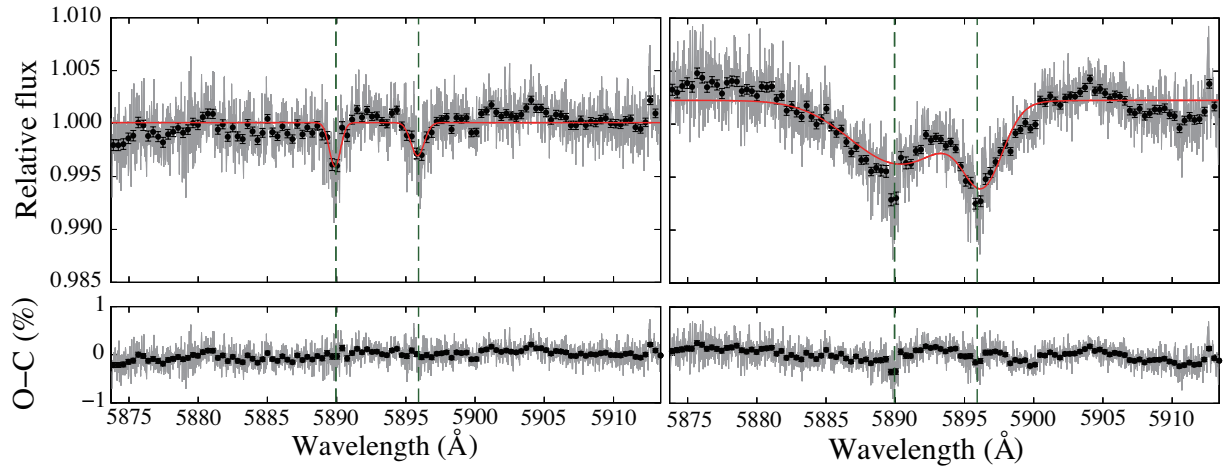}
\end{center}
\vspace{6mm}
\caption{Top panel shows the spectrum during transit divided by that before transit (left) and the spectrum during transit divided by that after transit (right). Red line indicates the fitting result obtained with a flat line and two Gaussians between 5873.7 and 5913.4 \mbox{\AA}. The bottom panel shows the residuals of fitting.}
\label{fig:resultWASP76_prv_only}
\end{figure*}

We found a wide trend around the Na absorption lines. To investigate the cause of this feature, we separated the template stellar spectrum into two types, namely spectra summed before and after transit, respectively. The left side of figure~\ref{fig:resultWASP76_prv_only} shows the transmission spectrum ($T_{\rm in/before}(\lambda)$) based on the template stellar spectrum summed before transit and the right side of figure~\ref{fig:resultWASP76_prv_only} shows the transmission spectrum ($T_{\rm in/after}(\lambda)$) based on by the template stellar spectrum summed after transit. The wide trend appears for only $T_{\rm in/after}(\lambda)$. Therefore, this trend results from the data after transit.

WASP-76 has a binary whose separation is 0".436 $\pm$ 0".003 and position angle is 215$^\circ$.9 $\pm$ 0$^\circ$.4 \citep{Bohn+2020}. From the image viewer at each observation time, it was found that part of the data during transit included this binary star. However, this did not cause the wide trend around the Na absorption lines because the time variation of this trend is different from the period when the binary was in the slit.

\begin {figure*} [htbp]
\begin{center}
 \includegraphics[width = 16.0cm]{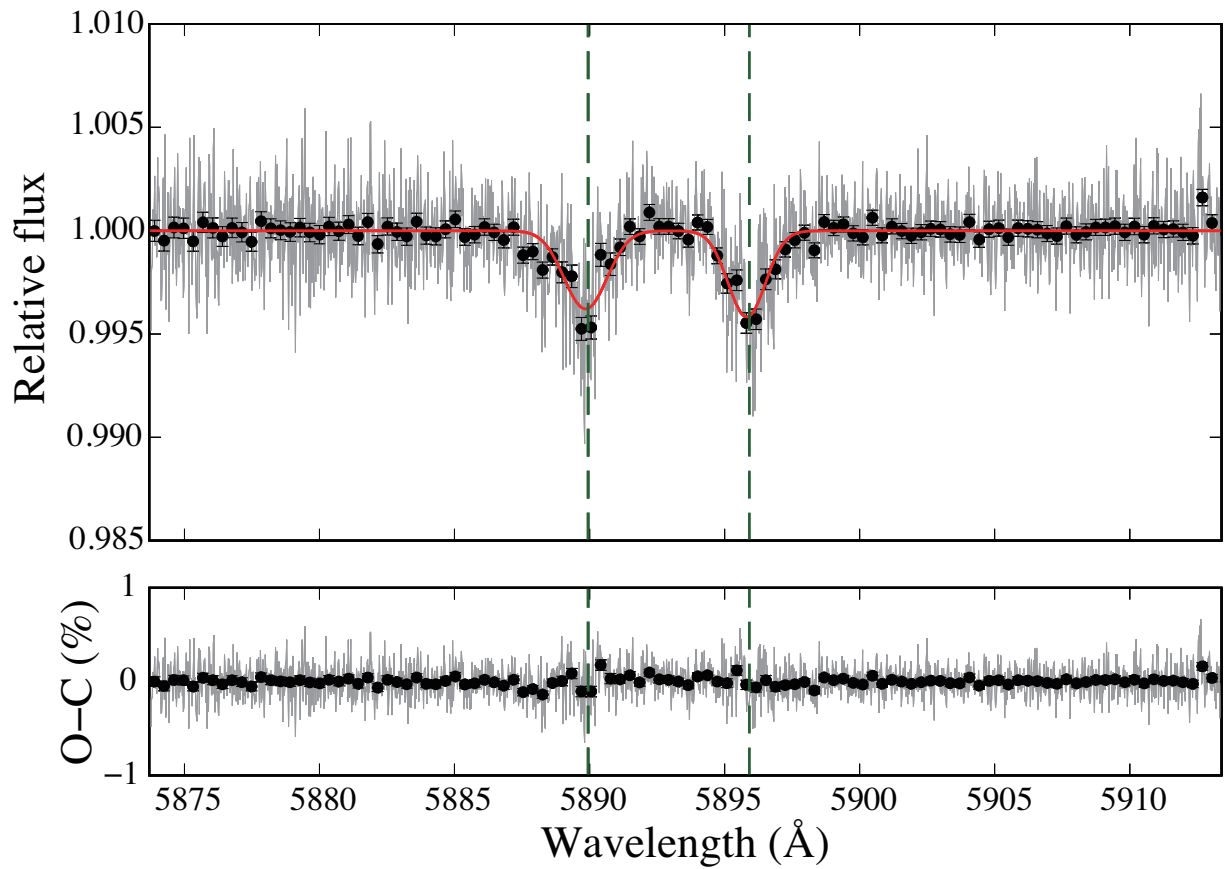}
\end{center}
\caption{Transmission spectrum after correction  {of a wide trend}. Red line indicates the fitting result obtained with a flat line and two Gaussians between 5873.7 and 5913.4 \mbox{\AA}. The bottom panel shows the residuals of fitting.}
\label{fig:resultWASP76_prvkai}
\end{figure*}

Because this trend is not the atmospheric signal, we corrected the variation with linear interpolation in each frame except around Na absorption lines (5887.22 -- 5897.98 \mbox{\AA}) to preserve the atmosphere's feature, and combined the transmission spectrum frames to create the transmission spectrum (figure~\ref{fig:resultWASP76_prvkai}).  Line contrasts of 0.42 $\pm$ 0.03 \% (D1) and 0.38 $\pm$ 0.04 \% (D2), FWHM values of 1.63 $\pm$ 0.13 \mbox{\AA} (D1) and 1.87 $\pm$ 0.22 \mbox{\AA} (D2), and EWs of (7.29 $\pm$ 1.43)$\times$10$^{-3}$ \mbox{\AA} (D1) and (7.56 $\pm$ 2.38)$\times$10$^{-3}$ \mbox{\AA} (D2) were obtained for this transmission spectrum. Our absorption lines are {wider than those reported in previous studies} (table~\ref{table:results}). 

\begin{table}[htbp]
\tbl{Results}{%
\begin{tabular}{ccccc}
\hline \hline
& \multicolumn{2}{c}{Our results}
  & \multicolumn{2}{c}{Seidel et al. 2019} \\
& D1 & D2 & D1 & D2  \\
\hline
Line contrast (\%) & 0.42 $\pm$ 0.03 & 0.38 $\pm$ 0.04 & 0.508 $\pm$ 0.053 & 0.373 $\pm$ 0.091 \\
FWHM (\mbox{\AA}) & 1.63 $\pm$ 0.13 & 1.87 $\pm$ 0.22 & 0.680 $\pm$ 0.128 & 0.619 $\pm$ 0.174 \\
\hline
& \multicolumn{4}{c}{{Tabernero et al. 2021}} \\
& {D1 (T1)} & {D2 (T1)} & {D1 (T2)} & {D2 (T2)} \\
\hline
Line contrast (\%) & {0.385 $\pm$ 0.051} & {0.449 $\pm$ 0.049} & {0.294 $\pm$ 0.042} & {0.246 $\pm$ 0.037} \\
FWHM (\mbox{\AA}) & {0.417 $\pm$ 0.065} & {0.483 $\pm$ 0.061} & {0.464 $\pm$ 0.081} & {0.623 $\pm$ 0.112} \\
\hline
\end{tabular}}\label{table:results}
\end{table}

Recently, it was reported that the Rossiter-McLaughlin (RM) effect and center-to-limb variation (CLV) affect the Na absorption estimated from high-resolution spectra \citep{Czesla+2015,Yan+2017,Borsa&Zannoni2018}.

The RM effect depends on the rotation velocity of the star and where the planet transits on the surface of the star. As WASP-76 b is in polar orbit \citep{Brown+2017} and WASP-76 is a slow rotator, the RM effect is small. The feature of the RM effect was not detected at the noise level of 0.01 \% \citep{Seidel+2019}. This means that the RM effect in the WASP-76 system is under the photon noise level in our results (continuum: $\sim$ 0.01 \%, line core: $\sim$ 0.04 \% per pixel) and can thus be ignored.

The CLV effect depends on the stellar temperature and the impact parameter of the planet.
As WASP-76 is an F-type star (T$_{\rm eff}$ = 6250 K), its CLV effect is less than that for earlier-type stars. According to \citet{Yan+2017}, the CLV effect is about 0.25 \% for an integral passband of 0.75 \mbox{\AA} because the impact parameter of WASP-76 b is 0.14.
This value is below the photon noise level in our results ($\sim$ 0.038) and thus the CLV effect can be ignored.

\subsection{Calculation of atmospheric absorption}

We detected the Na absorption lines of the planet atmosphere from the transmission spectrum of WASP-76 b. 

Based on \citet{Snellen+2008}, we estimated the planetary absorption signals at various passbands from the transmission spectrum and the variation in the planetary absorption signals at each time point (transmission light curve). First, we determined the center wavelength ($\lambda_{\rm 0}$) via a Gaussian fitting of the stellar Na absorption lines and obtained the average value ($F_{\rm mid}$), which is integrated by a passband from the determined line center ($\Delta \lambda$), for the difference spectra at each time point ($F_{\rm dif}(\lambda,t)$):
\begin{equation}
F_{\rm mid}(t) = \frac{\int_{\lambda_0-\frac{1}{2} \Delta \lambda}^{\lambda_0+\frac{1}{2} \Delta \lambda} F_{\rm dif}(\lambda,t) d\lambda}{\int_{\lambda_0-\frac{1}{2} \Delta \lambda}^{\lambda_0+\frac{1}{2} \Delta \lambda} d\lambda}.
\end{equation}
Then, we obtained the average of the integrated value in the continuum region on both sides of the Na absorption lines as 
\begin{eqnarray}
F_{\rm left}(t) = \frac{\int_{\rm A}^{\rm B} F_{\rm dif}(\lambda,t) d\lambda}{\int_{\rm A}^{\rm B} d\lambda}, \\ 
F_{\rm right}(t) =  \frac{\int_{\rm C}^{\rm D} F_{\rm dif}(\lambda,t) d\lambda}{\int_{\rm C}^{\rm D} d\lambda}.
\end{eqnarray}
\begin{center}
(A,B,C,D : suitably chosen coefficients)
\end{center}
Finally, by comparing $F_{\rm left}, F_{\rm right}$, and $F_{\rm mid}$, we obtained the planetary absorption signals of the Na D lines ($F_{\rm add}$) as 
\begin{equation}
F_{\rm add}(t)=  \frac{2 F_{\rm mid}(t)}{F_{\rm left}(t)+F_{\rm right}(t)}.
\end{equation}
Using this method, we also obtained the planetary absorption signals with various integral passbands for the transmission spectrum ($F_{\rm trans} (\lambda)$).

\begin {figure} [htbp]
\begin{center}
 \includegraphics[width = 8.0cm]{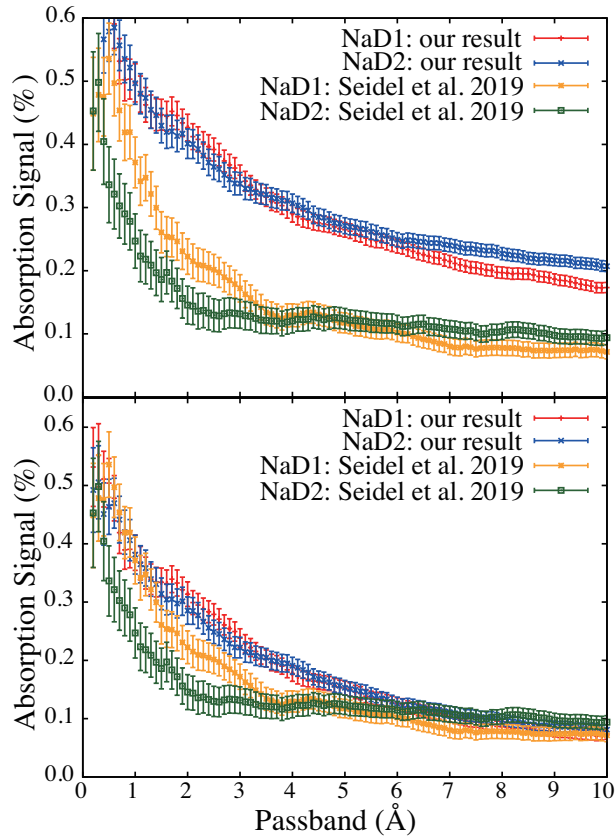}
\end{center}
\caption{Absorption of NaD1 (red) and D2 (blue) for each integral passband for our results before (top) and after (bottom) correction and for the results of \cite{Seidel+2019} (NaD1: orange, NaD2: green)} 
\label{fig:resultWASP76_tokahaba}
\end{figure}

By comparing the average of the integrated value in the left continuum region, which ranges from 5874.89 \mbox{\AA} to 5886.89 \mbox{\AA}, and that in the right continuum region, which ranges from 5898.89 \mbox{\AA} to 5910.89 \mbox{\AA}, we estimated the planetary absorption around the Na D lines for each passband (0.2--10 \mbox{\AA} in 0.1 \mbox{\AA} steps) for the original and corrected spectra and compared the results with those in previous work \citet{Seidel+2019} (figure~\ref{fig:resultWASP76_tokahaba}).
No differences were found between the Na D1 and NaD2 lines and the absorption signal for the corrected spectra is mostly consistent with that in the previous work.

\begin{figure*}[htbp]
 \begin{center}
 \includegraphics[width =16cm]{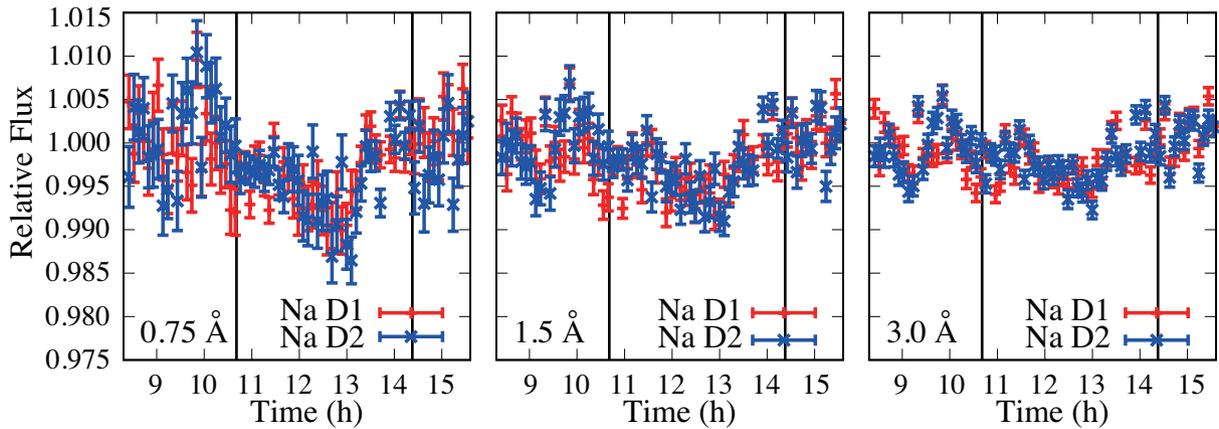}
  \end{center}
  \vspace{-6mm}
 \caption{Transmission light curves before correction integrated over passbands of 0.75 (left), 1.5 (center), and 3.0 (right) \mbox{\AA} around Na D1 (red) and Na D2 (blue) lines. The vertical lines represent the start (left) and end (right) of transit.}
 \label{fig:multi_zu}
\end{figure*}

We also created transmission light curves integrated over the passbands of 0.75, 1.5, and 3.0 \mbox{\AA} (figure~\ref{fig:multi_zu}) using the same continuum region.
The results show a significant reduction during transit for all passbands.

\section{Discussion}
\subsection{Effect of instrumental variation after correction}
Our results include instrumental variation and have a wide trend.
To confirm our correction of these variations, we used the empirical Monte Carlo (EMC) method as done by \citet{Redfield+2008}.

We created a template stellar spectrum ($F_{\rm star}$) and a master in-transit spectrum (that is, a normalized spectrum that combines all spectra during transit, $F_{\rm master-in}$) with randomly extracted frames from the frames of out-of-transit data only ($T_{\rm out-out}$) and the frames of in-transit data only ($T_{\rm in-in}$). We also  created  $F_{\rm star}$ with randomly extracted frames from the frames of out-of-transit data only and $F_{\rm master-in}$ with randomly extracted frames from the frames of in-transit data only ($T_{\rm in-out}$) . Note that the number of extracted frames is the same as that when we created the original spectra ($F_{\rm star}$: 35 frames, $F_{\rm master-in}$: 35 frames). We iterated the above process 20,000 times and examined the distribution of the absorption signal at each integral width ({3 and 6 \mbox{\AA}}).

\begin {figure*} [htbp]
\begin{center}
 \includegraphics[width = 16.0cm]{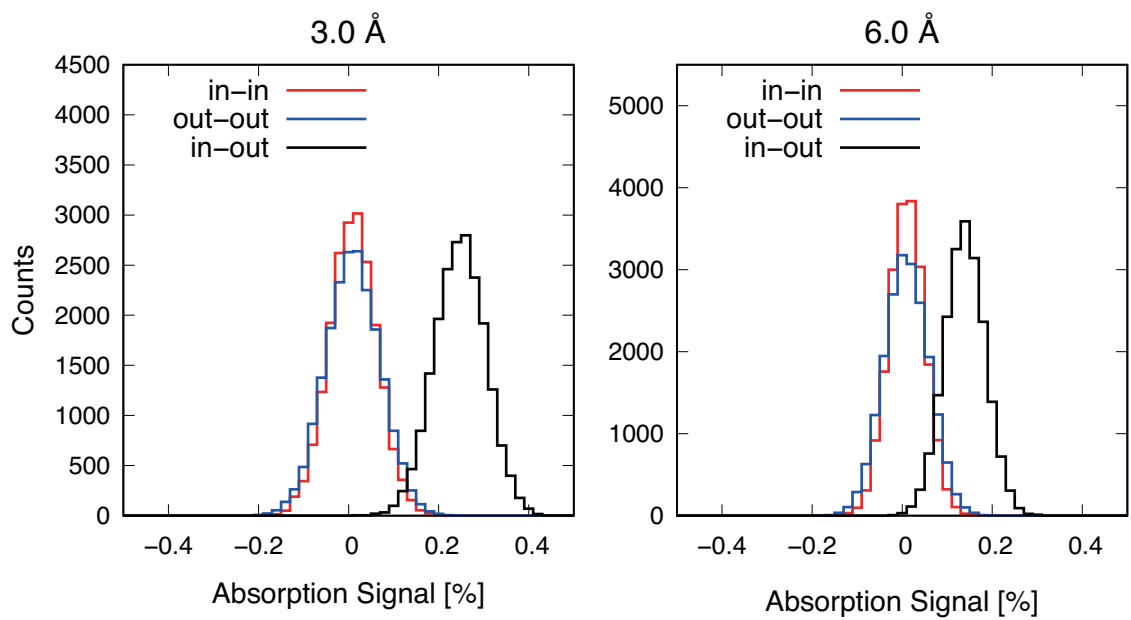}
\end{center}
\caption{Distribution of Absorption signal integrated over the passbands of  3.0 (left) and 6.0 (right) \mbox{\AA} around Na D lines.}
\label{fig:WASP76_histogram}
\end{figure*}

The absorption signals of $T_{\rm in-in}$, $T_{\rm out-out}$, and $T_{\rm in-out}$ are -0.000 $\pm$ 0.052 $\%$, -0.001 $\pm$ 0.060 $\%$, and 0.238 $\pm$ 0.057 $\%$ for 3 \mbox{\AA}, and 0.000 $\pm$ 0.040 $\%$, -0.001 $\pm$ 0.050 $\%$, and 0.129 $\pm$ 0.044 $\%$ for 6 \mbox{\AA} (figure~\ref{fig:WASP76_histogram}).
As the standard deviation of $T_{\rm out-out}$ includes the photon-noise and systematic error, we use it as the error of Na D absorption. Each integral passband for $T_{\rm in-out}$ and $F_{\rm trans} (\lambda)$ are different due to the planetary velocity shift, however its difference is within the error (right of figure~\ref{fig:WASP76_velocity}).
{The detection of Na absorption signals after correction is $\sim $ 4.0 $\sigma$ (3 \mbox{\AA}) and $\sim $ 2.6 $\sigma$ (6 \mbox{\AA}).}

\subsection{Comparison with theoretical spectra}
The shape of absorption lines is related to the temperature and state of the transmitted atmosphere. 
We first compare the observed Na absorption line with the theoretical spectra to obtain the relation between temperature and pressure in the atmosphere.

We calculated the theoretical transmission spectra using the model of \citet{2018ApJ...853....7K} basically in the same way as done by \citet{2014ApJ...790..108F} and \citet{Fukui:2016ky}. We calculated the thermochemical equilibrium abundances assuming solar elemental abundance ratios of \citet{2003ApJ...591.1220L}. The thermal ionization of Na was also considered. Line data of Na were taken from \citet{1992RMxAA..23...45K}.
We found an appropriate reference radius at 10~bar so that the average transit depth over the wavelength ranges of 5874.89 -- 5886.89~\mbox{\AA} and 5898.89 -- 5910.89~\mbox{\AA} matched the transit radius measured by \cite{Seidel+2019}, 2.078~$\rm R_J$, and the stellar radius revised by \cite{2018A&A...616A...1G}, 1.969~$\rm R_{\odot}$.
We have confirmed that the theoretical transmission spectrum calculated with our model can well reproduce that calculated with $^{\pi} \eta$ code \citep{2018A&A...612A..53P}, which was the one used by \citet{Seidel+2019}, when adopting the same parameters.

To explore the atmospheric thermal structure of WASP-76 b from the observed Na absorption line profiles, we considered both isothermal and non-isothermal atmospheres. For the isothermal case, we calculated the model spectra for the temperature range of 1000 -- 5500~K with a grid of 100~K. For the non-isothermal case, we used the analytical model of \cite{Guillot2010} (Eq.~29 in their paper), which is parameterized with the following five parameters: heat-redistribution factor $f$, intrinsic temperature $T_\mathrm{int}$, irradiation temperature $T_\mathrm{irr}$, and average opacity in the optical range $k_\mathrm{v}$ and infrared range $k_\mathrm{th}$. We assumed that $f = 0.25$, which corresponds to the case where the irradiation is effectively redistributed over the entire planetary surface, since we observed the transit. For $T_\mathrm{int}$, given that its effect on the temperature of the upper atmosphere is small, we fixed $T_\mathrm{int}$ to be 100~K. For $T_\mathrm{irr}$, we used the value for WASP-76 b (3119~K). For the other two parameters, we varied the values of $k_\mathrm{th}$ and $\gamma = k_\mathrm{v}/k_\mathrm{th}$ in the ranges of $10^{-8}$ -- $1$~$\mathrm{cm^2}$~$\mathrm{g}^{-1}$ and $10^{-2}$ -- $10^{2}$ with grids of 1 and 0.5 orders of magnitude, respectively.

For better accuracy, we converted wavelength space into velocity space around each Na absorption line and averaged the relative flux.
We defined the center wavelength of the Na D1 and D2 lines as 5895.92 and 5889.95 \mbox{\AA}, respectively, which were estimated from the fitting of the stellar spectra by a Gaussian.
In addition, to clarify the transmitted range at each wavelength, we converted the spectra into altitudes using 
\begin{equation}    
z (\lambda) = R_{p} \left( \sqrt{ 1+ \left( 1- \frac{R_s^2}{R_p^2} \right) \left( F_{\rm trans}(\lambda)-1 \right) } - 1 \right)
\end{equation}
where $R_{p}$ is the planet radius, $R_{s}$ is the stellar radius and $z (\lambda)$ and $F_{\rm trans}(\lambda)$ are the altitude and transmission spectra at each wavelength, respectively. The values of these parameters are shown in table~\ref{table:param}.
High resolution spectroscopy cannot obtain the absolute flux because the comparison stars cannot be observed simultaneously and the sky effect cannot be estimated (unlike for photometry and low resolution spectroscopy).
We calculated the relative altitude under the assumption that the reference range (the left and right continuum regions) is the planet radius.

\begin{figure}[htbp]	
 \begin{center}
 \includegraphics[width=16.0cm]{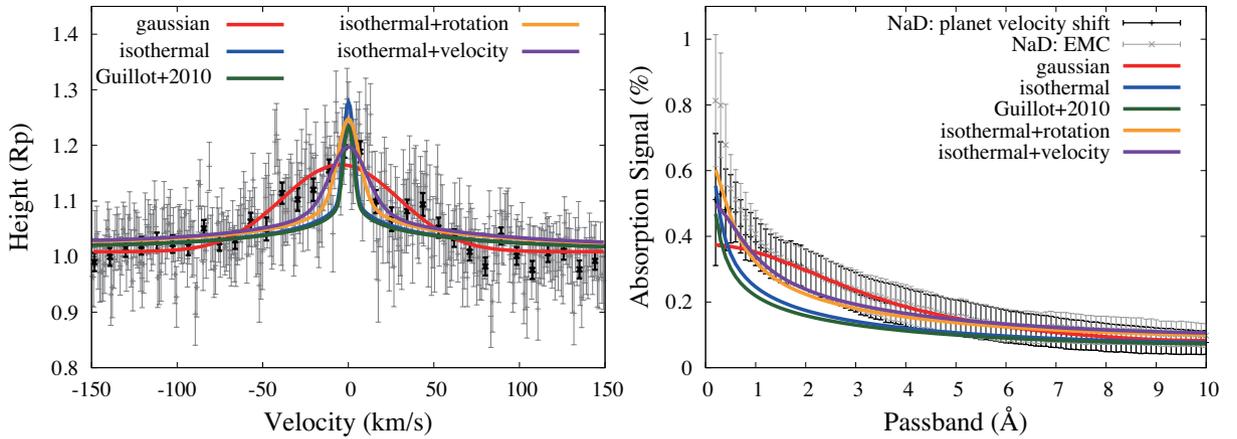}
\end{center}
\caption{The figure on the left is the average values of transmission spectra of Na D lines in velocity space. Our results are indicated in gray and binned by 10 pixels in black. The figure on the right is the average absorption of Na D lines. Black is obtained from equation (8) and gray is obtained from $T_{\rm in-out}$. Both of their errors are estimated from the standard deviation of $T_{\rm out-out}$. The red, blue, green, orange, and navy lines of both figures are the best fitted spectra by Gaussian, isothermal model, analytic radiative equilibrium model (Guillot+2010), isothermal model with rotation effect, and isothermal model with rotation velocity as free parameter.} 
\label{fig:WASP76_velocity}
\end{figure}

From comparisons with the isothermal model and an analytic radiative equilibrium model \citep{Guillot2010}, we found that the best fitting is achieved with the isothermal model for {3700 K} (blue line in figure~\ref{fig:WASP76_velocity}). Nevertheless, the broad width cannot be explained by this model. The rotation of WASP-76 b is about {6.0 km/s} at the planet radius under the assumption of tidal locking {, which is calculated using 2$\pi R_{\rm P} / P$ }. We included the rotation in the theoretical Na D absorption lines and compared them again. The isothermal model for {4200 K} (orange line in figure~\ref{fig:WASP76_velocity}) achieved the best fitting and had $\sim $ 1.1 times lower reduced chi-square than that of the best fitted isothermal model that did not consider the rotation (table~\ref{table:chi-squre}).
Furthermore, after fitting our data with the isothermal model, in which the rotation velocity was the free parameter, we found that the best fitted model was the isothermal model for {4300 K} with a rotation velocity of $\pm$ 11 km/s (purple line in figure~\ref{fig:WASP76_velocity}).

We also calculate the absorption for each integral passband for these best-fitted models and compared them with our observation result including the more accurate error obtained with the standard deviation of $T_{\rm out-out}$. 
{Our observation result is only consistent with the isothermal model considering the spread of velocity. Furthermore, the isothermal models with a velocity whose temperature is between 3900 K and 5000 K are also consistent with our result in the range of error on the integral passband (right of figure~\ref{fig:WASP76_velocity}; table~\ref{table:chi-squre}). }

\begin{figure}[htbp]	
 \begin{center}
 \includegraphics[width=16.0cm]{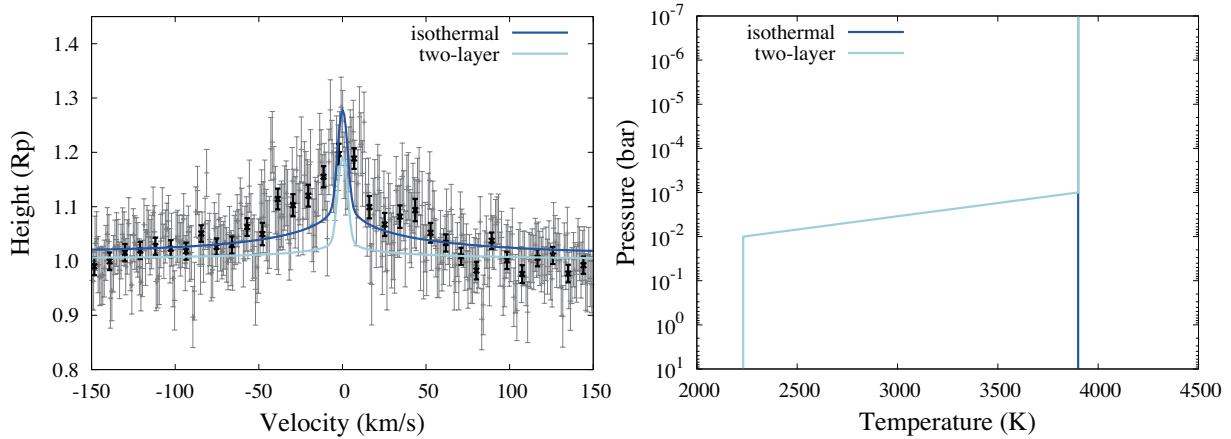}
\end{center}
\caption{Left: same as in figure~\ref{fig:WASP76_velocity} with an isothermal model (blue) and the model whose temperature is fixed under $10^{-3}$ bar (two-layer; light blue). Right: T-P profile for an isothermal model (blue) and a two-layer model (light blue). }
\label{fig:WASP76_two-layer}
\end{figure}

These results show that the value of a rotation velocity is necessary to fit both the transmission spectrum and absorption signal like recent studies with HARPS and ESPRESSO (e.g. \cite{Seidel+2020, Seidel+2021}) but difficult to be estimated without higher accurate observation data  {and} the atmosphere is hotter than the equilibrium temperature ($\sim$ 2160 K). 
Its temperature is hotter than the retrieved temperature from the low-resolution spectra obtained with HST but it is mostly consistent in the upper atmosphere with the model included the NUV and optical absorption by atoms and molecules such as TiO/VO (see Figure~7 in \cite{Edwards+2020}, Figure~2 in \cite{Lothringer+2020}). 
However, the temperature in the lower atmosphere is different from that obtained from the low-resolution spectra and the above models.
To create the best fitting model for our data assuming that the temperature in the lower atmosphere (under 10$^{-3}$ bar) is fixed the retrieved temperature from low-resolution spectra ($\sim$ 2230 K) , we should consider a model with a higher temperature in the upper atmosphere and with a larger additional velocity, because the absorption lines become sharper and shallower at lower temperatures in the lower atmosphere (figure~\ref{fig:WASP76_two-layer}).
To identify T-P profile more clearly, we should combine high-resolution spectroscopy, which is used mainly for observations of the upper atmosphere and low-resolution spectroscopy, which can be used for observing the lower atmosphere.

\begin{table}[htbp]
\tbl{Chi-square value for each fitting model}{
\scalebox{1.2}[1.2]{ 
\begin{tabular}{cccccc}
\hline \hline
& & \multicolumn{4}{c}{Transmission spectrum in velocity space} \\
 & & gaussian & isothermal & Guillot+2010 & isothermal \& rotation \\
 & & & {3700~K} & $k_{\rm th}$,$\gamma$ = {$10^{-7}$},$10^{1}$ & {4200~K} \\
\hline
& $\chi^2$ / dof & 247.8 / 323 & {332.1~/~327} & {341.7~/~326} &  {292.3~/~327} \\ 
& reduced-$\chi^2$ & 0.77 & {1.02} & {1.05} & {0.90} \\
\hline
& & \multicolumn{2}{c}{{\boldmath $\rm isothermal \ \& \ velocity$}} & \multicolumn{2}{c}{two-layer}\\
& & \multicolumn{2}{c}{{\boldmath $\rm 4300~K~\&~\pm~11~km/s$}} & \multicolumn{2}{c}{{3700~K} - 2230 K (fixed)}\\
\hline
& $\chi^2$ / dof &  \multicolumn{2}{c}{{\boldmath $\rm 280.7~/~326$}} &  \multicolumn{2}{c}{{511.0~/~327}} \\ 
& reduced-$\chi^2$ & \multicolumn{2}{c}{{\boldmath $\rm 0.86$}} & \multicolumn{2}{c}{{1.56}} \\
\hline
& & \multicolumn{4}{c}{Absorption signal for each passband} \\
 & & gaussian & isothermal & Guillot+2010 & isothermal \& rotation \\
 & & & {3700 K} & $k_{\rm th}$,$\gamma$ = {$10^{-7}$},$10^{1}$ & {4200 K} \\
\hline
$F_{add}$ & $\chi^2$ / dof & 6.9 / 91 & {120.9} / 98 &  {164.3} / 97 & {37.1} / 98 \\ 
& reduced-$\chi^2$ & 0.08 & {1.23} & {1.69} & {0.38} \\
EMC & $\chi^2$ / dof & 36.8 / 91 & {171.5} / 98 & {227.5} / 97 & {47.3} / 98 \\ 
& reduced-$\chi^2$ & 0.40 & {1.75} & {2.35} & {0.48} \\
\hline
& & \multicolumn{4}{c}{{{isothermal \ \& \ velocity}}} \\ 
& & \multicolumn{2}{c}{{{\boldmath $\rm 4300 \ K \ \& \ \pm 11 \ km/s $}}} & {3900 K \ \& \ $\pm$ 0 km/s} & {5000 K \ \& \ $\pm$ 37 km/s} \\
\hline
$F_{add}$ & $\chi^2$ / dof & \multicolumn{2}{c}{{{\boldmath $\rm 30.0 \ / \ 97 $}}} & {87.3 / 97} & {129.8 / 97} \\ 
& reduced-$\chi^2$ & \multicolumn{2}{c}{{{\boldmath $\rm 0.31$}}} & {0.90} & {1.33} \\
EMC & $\chi^2$ / dof & \multicolumn{2}{c}{{{\boldmath $\rm 35.8 \ / \ 97 $}}} & {124.0 / 97} & {96.6 / 97} \\ 
& reduced-$\chi^2$ & \multicolumn{2}{c}{{{\boldmath $\rm 0.37$ }}} & {1.28} & {1.00} \\
\hline
\end{tabular}}}\label{table:chi-squre}
\begin{flushleft}
*dof = degree of freedom
\end{flushleft}
\end{table}

In addition, to estimate the spread of the line width, we fit our Na absorption lines by a Gaussian in velocity space (red line in figure~\ref{fig:WASP76_velocity}). The center is blue-shifted by about 4.5 km/s and the FWHM value is 80.7 $\pm$ 7.9 km/s, which is about 3.0 times broader than that in previous work and about 20 times broader than the sonic speed.

The spread of the width suggests the existence of additional wind (e.g., super-rotation, day-to-night side wind) other than the wind caused by rotation \citep{Seidel+2020} and time variation because the observation days in previous work and this work are different. There is also possibility that there is an instrumental difference. The sodium velocity is up to $\sim$ 100 km/s considering charge exchange in the sodium exosphere at Jupiter or pickup ions in the Io plasma torus \citep{Gebek&Oza2020}. There is thus a possibility that this spread of the width can be explained by the plasma-driven escape of Na at exo-Io or exo-torus.

Our average transmission spectrum of Na D lines is slightly asymmetric and blue-shifted. This supports the day-night chemistry from the detected iron absorption lines with high-resolution spectroscopy \citep{Ehrenreich+2020}.  

\section{Summary}
 We observed the ultra-hot Jupiter WASP-76 b on UT September 10, 2018, with Subaru/HDS to detect the Na D absorption lines for the atmosphere. 
The observed data have telluric absorption lines, stellar absorption lines, and instrumental profile in addition to the absorption lines of the exoplanetary atmosphere.
We carefully removed the unnecessary spectra and instrumental profile from the Subaru/HDS data and combined the transmission spectra during transit considering the planet radial velocity.

Na D lines were detected and fitted after correcting for the wide trend around the Na absorption lines with a flat line and two Gaussians. We obtained line contrasts of 0.42 $\pm$ 0.03 \% (D1) and 0.38 $\pm$ 0.04 \% (D2), FWHM values of 1.63 $\pm$ 0.13 \mbox{\AA} (D1) and 1.87 $\pm$ 0.22 \mbox{\AA} (D2), and EWs of (7.29 $\pm$ 1.43) $\times$ 10$^{-3}$ \mbox{\AA} (D1) and (7.56 $\pm$ 2.38) $\times$ 10$^{-3}$ \mbox{\AA} (D2).
These Na D absorption lines were shallower and broader than those in previous work, but the absorption signals over a given passband are consistent with those in previous work. 
Our results also show that the spread of the width is 80.7 $\pm$ 7.9 km/s which is about 3.0 times broader than previous work in velocity space.
This result suggests the existence of additional wind. To identify this trend, we should compare our results with a more sophisticated model considering various additional wind patterns, such as that by \citet{Seidel+2020}.

In addition, Na D lines were compared with the theoretical transmission spectra calculated based on an isothermal model and an analytic radiative equilibrium model with and without rotation. 
Isothermal models whose temperature was 3700 K (without rotation), and 4200 K (with rotation) were estimated to have the best fit. 
These results indicate that the atmospheric temperature is higher than the equilibrium temperature ($\sim$ 2160 K) and mostly consistent with the temperature in the upper atmosphere estimated by the low-resolution spectral data obtained with HST at NUV, optical and near-infrared wavelengths. 
In the future, we will investigate the temperature structure in more detail by combining high- and low-resolution spectra. 

\begin{ack}
We would like to thank L.~Pino for fruitful discussion and kindly providing us his calculation data for the model comparison, J.~V.~Seidel for providing observation data, and A.~Tajitsu for the support of our Subaru/HDS observation.
We also thank MuSCAT team especially M.~Mori who teaches how to use the software "emcee".
We are grateful to Forte (www.forte-science.co.jp) for English editing.
This research is based on data collected at Subaru Telescope,
which is operated by the National Astronomical Observatory of Japan.
We are honored and grateful for the opportunity of observing the Universe from Maunakea, which has the cultural, historical and natural significance in Hawaii.
The part of our data analysis was carried out on common use data analysis computer system at the Astronomy Data Center, ADC, of the National Astronomical Observatory of Japan.
This work is partly supported by JSPS KAKENHI Grant Number JP21K13955, JP18H05439, JST PRESTO Grant Number JPMJPR1775, the Astrobiology Center of National Institutes of Natural Sciences (NINS) (Grant Number AB031010).
Y.K. is supported by Special Postdoctoral Researcher Program at RIKEN.
\end{ack}

\bibliographystyle{apj}
\bibliography{reference}

\begin{thebibliography}{64}
\expandafter\ifx\csname natexlab\endcsname\relax\def\natexlab#1{#1}\fi

\bibitem[{{Astudillo-Defru} \& {Rojo}(2013)}]{Astudillo-Defru&Rojo2013}
{Astudillo-Defru}, N., \& {Rojo}, P. 2013, \aap, 557, A56

\bibitem[{{Barman} {et~al.}(2001){Barman}, {Hauschildt}, \&
  {Allard}}]{Barman+2001}
{Barman}, T.~S., {Hauschildt}, P.~H., \& {Allard}, F. 2001, \apj, 556, 885

\bibitem[{{Bohn} {et~al.}(2020){Bohn}, {Southworth}, {Ginski}, {Kenworthy},
  {Maxted}, \& {Evans}}]{Bohn+2020}
{Bohn}, A.~J., {Southworth}, J., {Ginski}, C., {et~al.} 2020, \aap, 635, A73

\bibitem[{{Borsa} \& {Zannoni}(2018)}]{Borsa&Zannoni2018}
{Borsa}, F., \& {Zannoni}, A. 2018, \aap, 617, A134

\bibitem[{{Brown} {et~al.}(2017){Brown}, {Triaud}, {Doyle}, {Gillon}, {Lendl},
  {Anderson}, {Collier Cameron}, {H{\'e}brard}, {Hellier}, {Lovis}, {Maxted},
  {Pepe}, {Pollacco}, {Queloz}, \& {Smalley}}]{Brown+2017}
{Brown}, D.~J.~A., {Triaud}, A.~H.~M.~J., {Doyle}, A.~P., {et~al.} 2017,
  \mnras, 464, 810

\bibitem[{{Brown}(2001)}]{Brown+2001}
{Brown}, T.~M. 2001, \apj, 553, 1006

\bibitem[{{Burton} {et~al.}(2015){Burton}, {Watson}, {Rodr{\'\i}guez-Gil},
  {Skillen}, {Littlefair}, {Dhillon}, \& {Pollacco}}]{Burton+2015}
{Burton}, J.~R., {Watson}, C.~A., {Rodr{\'\i}guez-Gil}, P., {et~al.} 2015,
  \mnras, 446, 1071

\bibitem[{{Casasayas-Barris} {et~al.}(2017){Casasayas-Barris}, {Palle},
  {Nowak}, {Yan}, {Nortmann}, \& {Murgas}}]{Casasayas-Barris+2017}
{Casasayas-Barris}, N., {Palle}, E., {Nowak}, G., {et~al.} 2017, \aap, 608,
  A135

\bibitem[{{Casasayas-Barris} {et~al.}(2018){Casasayas-Barris}, {Pall{\'e}},
  {Yan}, {Chen}, {Albrecht}, {Nortmann}, {Van Eylen}, {Snellen}, {Talens},
  {Gonz{\'a}lez Hern{\'a}ndez}, {Rebolo}, \& {Otten}}]{Casasayas-Barris+2018}
{Casasayas-Barris}, N., {Pall{\'e}}, E., {Yan}, F., {et~al.} 2018, \aap, 616,
  A151

\bibitem[{{Chen} {et~al.}(2020){Chen}, {Casasayas-Barris}, {Pall{\'e}},
  {Welbanks}, {Madhusudhan}, {Luque}, \& {Murgas}}]{Chen+2020}
{Chen}, G., {Casasayas-Barris}, N., {Pall{\'e}}, E., {et~al.} 2020, \aap, 642,
  A54

\bibitem[{{Czesla} {et~al.}(2015){Czesla}, {Klocov{\'a}}, {Khalafinejad},
  {Wolter}, \& {Schmitt}}]{Czesla+2015}
{Czesla}, S., {Klocov{\'a}}, T., {Khalafinejad}, S., {Wolter}, U., \&
  {Schmitt}, J.~H.~M.~M. 2015, \aap, 582, A51

\bibitem[{{Dempsey} {et~al.}(2013){Dempsey}, {Friberg}, {Jenness}, {Tilanus},
  {Thomas}, {Holland}, {Bintley}, {Berry}, {Chapin}, {Chrysostomou}, {Davis},
  {Gibb}, {Parsons}, \& {Robson}}]{Dempsey+2013}
{Dempsey}, J.~T., {Friberg}, P., {Jenness}, T., {et~al.} 2013, \mnras, 430,
  2534

\bibitem[{{Edwards} {et~al.}(2020){Edwards}, {Changeat}, {Baeyens}, {Tsiaras},
  {Al-Refaie}, {Taylor}, {Yip}, {Bieger}, {Blain}, {Gressier}, {Guilluy},
  {Jaziri}, {Kiefer}, {Modirrousta-Galian}, {Morvan}, {Mugnai}, {Pluriel},
  {Poveda}, {Skaf}, {Whiteford}, {Wright}, {Zingales}, {Charnay}, {Drossart},
  {Leconte}, {Venot}, {Waldmann}, \& {Beaulieu}}]{Edwards+2020}
{Edwards}, B., {Changeat}, Q., {Baeyens}, R., {et~al.} 2020, \aj, 160, 8

\bibitem[{{Ehrenreich} {et~al.}(2020){Ehrenreich}, {Lovis}, {Allart}, {Zapatero
  Osorio}, {Pepe}, {Cristiani}, {Rebolo}, {Santos}, {Borsa}, {Demangeon},
  {Dumusque}, {Gonz{\'a}lez Hern{\'a}ndez}, {Casasayas-Barris},
  {S{\'e}gransan}, {Sousa}, {Abreu}, {Adibekyan}, {Affolter}, {Allende Prieto},
  {Alibert}, {Aliverti}, {Alves}, {Amate}, {Avila}, {Baldini}, {Bandy}, {Benz},
  {Bianco}, {Bolmont}, {Bouchy}, {Bourrier}, {Broeg}, {Cabral}, {Calderone},
  {Pall{\'e}}, {Cegla}, {Cirami}, {Coelho}, {Conconi}, {Coretti}, {Cumani},
  {Cupani}, {Dekker}, {Delabre}, {Deiries}, {D'Odorico}, {Di Marcantonio},
  {Figueira}, {Fragoso}, {Genolet}, {Genoni}, {G{\'e}nova Santos}, {Hara},
  {Hughes}, {Iwert}, {Kerber}, {Knudstrup}, {Land oni}, {Lavie}, {Lizon},
  {Lendl}, {Lo Curto}, {Maire}, {Manescau}, {Martins}, {M{\'e}gevand },
  {Mehner}, {Micela}, {Modigliani}, {Molaro}, {Monteiro}, {Monteiro},
  {Moschetti}, {M{\"u}ller}, {Nunes}, {Oggioni}, {Oliveira}, {Pariani},
  {Pasquini}, {Poretti}, {Rasilla}, {Redaelli}, {Riva}, {Santana Tschudi},
  {Santin}, {Santos}, {Segovia Milla}, {Seidel}, {Sosnowska}, {Sozzetti},
  {Span{\`o}}, {Su{\'a}rez Mascare{\~n}o}, {Tabernero}, {Tenegi}, {Udry},
  {Zanutta}, \& {Zerbi}}]{Ehrenreich+2020}
{Ehrenreich}, D., {Lovis}, C., {Allart}, R., {et~al.} 2020, \nat, 580, 597

\bibitem[{{Evans} {et~al.}(2016){Evans}, {Sing}, {Wakeford}, {Nikolov},
  {Ballester}, {Drummond}, {Kataria}, {Gibson}, {Amundsen}, \&
  {Spake}}]{Evans+2016}
{Evans}, T.~M., {Sing}, D.~K., {Wakeford}, H.~R., {et~al.} 2016, \apjl, 822, L4

\bibitem[{{Foreman-Mackey} {et~al.}(2013){Foreman-Mackey}, {Hogg}, {Lang}, \&
  {Goodman}}]{Foreman-Mackey+2013}
{Foreman-Mackey}, D., {Hogg}, D.~W., {Lang}, D., \& {Goodman}, J. 2013, \pasp,
  125, 306

\bibitem[{{Fortney} {et~al.}(2008){Fortney}, {Lodders}, {Marley}, \&
  {Freedman}}]{Fortney+2008}
{Fortney}, J.~J., {Lodders}, K., {Marley}, M.~S., \& {Freedman}, R.~S. 2008,
  \apj, 678, 1419

\bibitem[{{Fortney} {et~al.}(2010){Fortney}, {Shabram}, {Showman}, {Lian},
  {Freedman}, {Marley}, \& {Lewis}}]{Fortney+2010}
{Fortney}, J.~J., {Shabram}, M., {Showman}, A.~P., {et~al.} 2010, \apj, 709,
  1396

\bibitem[{{Fu} {et~al.}(2020){Fu}, {Deming}, {Lothringer}, {Nikolov}, {Sing},
  {Kempton}, {Ih}, {Evans}, {Stevenson}, {Wakeford}, {Rodriguez}, {Eastman},
  {Stassun}, {Henry}, {L{\'o}pez-Morales}, {Lendl}, {Conti}, {Stockdale},
  {Collins}, {Kielkopf}, {Barstow}, {Sanz-Forcada}, {Ehrenreich}, \&
  {Bourrier}}]{Fu+2020arXiv}
{Fu}, G., {Deming}, D., {Lothringer}, J., {et~al.} 2020, arXiv e-prints,
  arXiv:2005.02568

\bibitem[{{Fukui} {et~al.}(2014){Fukui}, {Kawashima}, {Ikoma}, {Narita},
  {Onitsuka}, {Ita}, {Onozato}, {Nishiyama}, {Baba}, {Ryu}, {Hirano}, {Hori},
  {Kurosaki}, {Kawauchi}, {Takahashi}, {Nagayama}, {Tamura}, {Kawai}, {Kuroda},
  {Nagayama}, {Ohta}, {Shimizu}, {Yanagisawa}, {Yoshida}, \&
  {Izumiura}}]{2014ApJ...790..108F}
{Fukui}, A., {Kawashima}, Y., {Ikoma}, M., {et~al.} 2014, \apj, 790, 108

\bibitem[{{Fukui} {et~al.}(2016){Fukui}, {Narita}, {Kawashima}, {Kusakabe},
  {Onitsuka}, {Ryu}, {Ikoma}, {Yanagisawa}, \& {Izumiura}}]{Fukui:2016ky}
{Fukui}, A., {Narita}, N., {Kawashima}, Y., {et~al.} 2016, \apj, 819, 27

\bibitem[{{Gaia Collaboration} {et~al.}(2018){Gaia Collaboration}, {Brown},
  {Vallenari}, {Prusti}, {de Bruijne}, {Babusiaux}, {Bailer-Jones}, {Biermann},
  {Evans}, {Eyer}, {Jansen}, {Jordi}, {Klioner}, {Lammers}, {Lindegren},
  {Luri}, {Mignard}, {Panem}, {Pourbaix}, {Randich}, {Sartoretti}, {Siddiqui},
  {Soubiran}, {van Leeuwen}, {Walton}, {Arenou}, {Bastian}, {Cropper},
  {Drimmel}, {Katz}, {Lattanzi}, {Bakker}, {Cacciari}, {Casta{\~n}eda},
  {Chaoul}, {Cheek}, {De Angeli}, {Fabricius}, {Guerra}, {Holl}, {Masana},
  {Messineo}, {Mowlavi}, {Nienartowicz}, {Panuzzo}, {Portell}, {Riello},
  {Seabroke}, {Tanga}, {Th{\'e}venin}, {Gracia-Abril}, {Comoretto},
  {Garcia-Reinaldos}, {Teyssier}, {Altmann}, {Andrae}, {Audard},
  {Bellas-Velidis}, {Benson}, {Berthier}, {Blomme}, {Burgess}, {Busso},
  {Carry}, {Cellino}, {Clementini}, {Clotet}, {Creevey}, {Davidson}, {De
  Ridder}, {Delchambre}, {Dell'Oro}, {Ducourant},
  {Fern{\'a}ndez-Hern{\'a}ndez}, {Fouesneau}, {Fr{\'e}mat}, {Galluccio},
  {Garc{\'\i}a-Torres}, {Gonz{\'a}lez-N{\'u}{\~n}ez}, {Gonz{\'a}lez-Vidal},
  {Gosset}, {Guy}, {Halbwachs}, {Hambly}, {Harrison}, {Hern{\'a}ndez},
  {Hestroffer}, {Hodgkin}, {Hutton}, {Jasniewicz}, {Jean-Antoine-Piccolo},
  {Jordan}, {Korn}, {Krone-Martins}, {Lanzafame}, {Lebzelter}, {L{\"o}ffler},
  {Manteiga}, {Marrese}, {Mart{\'\i}n-Fleitas}, {Moitinho}, {Mora}, {Muinonen},
  {Osinde}, {Pancino}, {Pauwels}, {Petit}, {Recio-Blanco}, {Richards},
  {Rimoldini}, {Robin}, {Sarro}, {Siopis}, {Smith}, {Sozzetti}, {S{\"u}veges},
  {Torra}, {van Reeven}, {Abbas}, {Abreu Aramburu}, {Accart}, {Aerts},
  {Altavilla}, {{\'A}lvarez}, {Alvarez}, {Alves}, {Anderson}, {Andrei},
  {Anglada Varela}, {Antiche}, {Antoja}, {Arcay}, {Astraatmadja}, {Bach},
  {Baker}, {Balaguer-N{\'u}{\~n}ez}, {Balm}, {Barache}, {Barata}, {Barbato},
  {Barblan}, {Barklem}, {Barrado}, {Barros}, {Barstow}, {Bartholom{\'e}
  Mu{\~n}oz}, {Bassilana}, {Becciani}, {Bellazzini}, {Berihuete}, {Bertone},
  {Bianchi}, {Bienaym{\'e}}, {Blanco-Cuaresma}, {Boch}, {Boeche}, {Bombrun},
  {Borrachero}, {Bossini}, {Bouquillon}, {Bourda}, {Bragaglia}, {Bramante},
  {Breddels}, {Bressan}, {Brouillet}, {Br{\"u}semeister}, {Brugaletta},
  {Bucciarelli}, {Burlacu}, {Busonero}, {Butkevich}, {Buzzi}, {Caffau},
  {Cancelliere}, {Cannizzaro}, {Cantat-Gaudin}, {Carballo}, {Carlucci},
  {Carrasco}, {Casamiquela}, {Castellani}, {Castro-Ginard}, {Charlot},
  {Chemin}, {Chiavassa}, {Cocozza}, {Costigan}, {Cowell}, {Crifo}, {Crosta},
  {Crowley}, {Cuypers}, {Dafonte}, {Damerdji}, {Dapergolas}, {David}, {David},
  {de Laverny}, {De Luise}, {De March}, {de Martino}, {de Souza}, {de Torres},
  {Debosscher}, {del Pozo}, {Delbo}, {Delgado}, {Delgado}, {Di Matteo},
  {Diakite}, {Diener}, {Distefano}, {Dolding}, {Drazinos}, {Dur{\'a}n},
  {Edvardsson}, {Enke}, {Eriksson}, {Esquej}, {Eynard Bontemps}, {Fabre},
  {Fabrizio}, {Faigler}, {Falc{\~a}o}, {Farr{\`a}s Casas}, {Federici},
  {Fedorets}, {Fernique}, {Figueras}, {Filippi}, {Findeisen}, {Fonti},
  {Fraile}, {Fraser}, {Fr{\'e}zouls}, {Gai}, {Galleti}, {Garabato},
  {Garc{\'\i}a-Sedano}, {Garofalo}, {Garralda}, {Gavel}, {Gavras}, {Gerssen},
  {Geyer}, {Giacobbe}, {Gilmore}, {Girona}, {Giuffrida}, {Glass}, {Gomes},
  {Granvik}, {Gueguen}, {Guerrier}, {Guiraud}, {Guti{\'e}rrez-S{\'a}nchez},
  {Haigron}, {Hatzidimitriou}, {Hauser}, {Haywood}, {Heiter}, {Helmi}, {Heu},
  {Hilger}, {Hobbs}, {Hofmann}, {Holland}, {Huckle}, {Hypki}, {Icardi},
  {Jan{\ss}en}, {Jevardat de Fombelle}, {Jonker}, {Juh{\'a}sz}, {Julbe},
  {Karampelas}, {Kewley}, {Klar}, {Kochoska}, {Kohley}, {Kolenberg},
  {Kontizas}, {Kontizas}, {Koposov}, {Kordopatis}, {Kostrzewa-Rutkowska},
  {Koubsky}, {Lambert}, {Lanza}, {Lasne}, {Lavigne}, {Le Fustec}, {Le
  Poncin-Lafitte}, {Lebreton}, {Leccia}, {Leclerc}, {Lecoeur-Taibi},
  {Lenhardt}, {Leroux}, {Liao}, {Licata}, {Lindstr{\o}m}, {Lister}, {Livanou},
  {Lobel}, {L{\'o}pez}, {Managau}, {Mann}, {Mantelet}, {Marchal}, {Marchant},
  {Marconi}, {Marinoni}, {Marschalk{\'o}}, {Marshall}, {Martino}, {Marton},
  {Mary}, {Massari}, {Matijevi{\v{c}}}, {Mazeh}, {McMillan}, {Messina},
  {Michalik}, {Millar}, {Molina}, {Molinaro}, {Moln{\'a}r}, {Montegriffo},
  {Mor}, {Morbidelli}, {Morel}, {Morris}, {Mulone}, {Muraveva}, {Musella},
  {Nelemans}, {Nicastro}, {Noval}, {O'Mullane}, {Ord{\'e}novic},
  {Ord{\'o}{\~n}ez-Blanco}, {Osborne}, {Pagani}, {Pagano}, {Pailler},
  {Palacin}, {Palaversa}, {Panahi}, {Pawlak}, {Piersimoni}, {Pineau}, {Plachy},
  {Plum}, {Poggio}, {Poujoulet}, {Pr{\v{s}}a}, {Pulone}, {Racero}, {Ragaini},
  {Rambaux}, {Ramos-Lerate}, {Regibo}, {Reyl{\'e}}, {Riclet}, {Ripepi}, {Riva},
  {Rivard}, {Rixon}, {Roegiers}, {Roelens}, {Romero-G{\'o}mez}, {Rowell},
  {Royer}, {Ruiz-Dern}, {Sadowski}, {Sagrist{\`a} Sell{\'e}s}, {Sahlmann},
  {Salgado}, {Salguero}, {Sanna}, {Santana-Ros}, {Sarasso}, {Savietto},
  {Schultheis}, {Sciacca}, {Segol}, {Segovia}, {S{\'e}gransan}, {Shih},
  {Siltala}, {Silva}, {Smart}, {Smith}, {Solano}, {Solitro}, {Sordo}, {Soria
  Nieto}, {Souchay}, {Spagna}, {Spoto}, {Stampa}, {Steele},
  {Steidelm{\"u}ller}, {Stephenson}, {Stoev}, {Suess}, {Surdej}, {Szabados},
  {Szegedi-Elek}, {Tapiador}, {Taris}, {Tauran}, {Taylor}, {Teixeira},
  {Terrett}, {Teyssand ier}, {Thuillot}, {Titarenko}, {Torra Clotet}, {Turon},
  {Ulla}, {Utrilla}, {Uzzi}, {Vaillant}, {Valentini}, {Valette}, {van Elteren},
  {Van Hemelryck}, {van Leeuwen}, {Vaschetto}, {Vecchiato}, {Veljanoski},
  {Viala}, {Vicente}, {Vogt}, {von Essen}, {Voss}, {Votruba}, {Voutsinas},
  {Walmsley}, {Weiler}, {Wertz}, {Wevers}, {Wyrzykowski}, {Yoldas},
  {{\v{Z}}erjal}, {Ziaeepour}, {Zorec}, {Zschocke}, {Zucker}, {Zurbach}, \&
  {Zwitter}}]{2018A&A...616A...1G}
{Gaia Collaboration}, {Brown}, A.~G.~A., {Vallenari}, A., {et~al.} 2018, \aap,
  616, A1

\bibitem[{{Gebek} \& {Oza}(2020)}]{Gebek&Oza2020}
{Gebek}, A., \& {Oza}, A.~V. 2020, arXiv e-prints, arXiv:2005.02536

\bibitem[{{Glebocki} {et~al.}(2000){Glebocki}, {Gnacinski}, \&
  {Stawikowski}}]{Glebocki+2000}
{Glebocki}, R., {Gnacinski}, P., \& {Stawikowski}, A. 2000, Acta Astronomica,
  50, 509

\bibitem[{{Guillot}(2010)}]{Guillot2010}
{Guillot}, T. 2010, \aap, 520, A27

\bibitem[{{Hoeijmakers} {et~al.}(2019){Hoeijmakers}, {Ehrenreich}, {Kitzmann},
  {Allart}, {Grimm}, {Seidel}, {Wyttenbach}, {Pino}, {Nielsen}, {Fisher},
  {Rimmer}, {Bourrier}, {Cegla}, {Lavie}, {Lovis}, {Patzer}, {Stock}, {Pepe},
  \& {Heng}}]{Hoeijmakers+2019}
{Hoeijmakers}, H.~J., {Ehrenreich}, D., {Kitzmann}, D., {et~al.} 2019, \aap,
  627, A165

\bibitem[{{Hoeijmakers} {et~al.}(2020){Hoeijmakers}, {Seidel}, {Pino},
  {Kitzmann}, {Sindel}, {Ehrenreich}, {Oza}, {Bourrier}, {Allart}, {Gebek},
  {Lovis}, {Yurchenko}, {Astudillo-Defru}, {Bayliss}, {Cegla}, {Lavie},
  {Lendl}, {Melo}, {Murgas}, {Nascimbeni}, {Pepe}, {S{\'e}gransan}, {Udry},
  {Wyttenbach}, \& {Heng}}]{Hoeijmakers+2020}
{Hoeijmakers}, H.~J., {Seidel}, J.~V., {Pino}, L., {et~al.} 2020, \aap, 641,
  A123

\bibitem[{{Hubbard} {et~al.}(2001){Hubbard}, {Fortney}, {Lunine}, {Burrows},
  {Sudarsky}, \& {Pinto}}]{Hubbard+2001}
{Hubbard}, W.~B., {Fortney}, J.~J., {Lunine}, J.~I., {et~al.} 2001, \apj, 560,
  413

\bibitem[{{Jensen} {et~al.}(2018){Jensen}, {Cauley}, {Redfield}, {Cochran}, \&
  {Endl}}]{Jensen+2018}
{Jensen}, A.~G., {Cauley}, P.~W., {Redfield}, S., {Cochran}, W.~D., \& {Endl},
  M. 2018, \aj, 156, 154

\bibitem[{{Kausch} {et~al.}(2015){Kausch}, {Noll}, {Smette}, {Kimeswenger},
  {Barden}, {Szyszka}, {Jones}, {Sana}, {Horst}, \& {Kerber}}]{Kausch+2015}
{Kausch}, W., {Noll}, S., {Smette}, A., {et~al.} 2015, \aap, 576, A78

\bibitem[{{Kawashima} \& {Ikoma}(2018)}]{2018ApJ...853....7K}
{Kawashima}, Y., \& {Ikoma}, M. 2018, \apj, 853, 7

\bibitem[{{Kawauchi} {et~al.}(2018){Kawauchi}, {Narita}, {Sato}, {Hirano},
  {Kawashima}, {Nakamoto}, {Yamashita}, \& {Tamura}}]{Kawauchi+2018}
{Kawauchi}, K., {Narita}, N., {Sato}, B., {et~al.} 2018, \pasj, 70, 84

\bibitem[{{Khalafinejad} {et~al.}(2017){Khalafinejad}, {von Essen},
  {Hoeijmakers}, {Zhou}, {Klocov{\'a}}, {Schmitt}, {Dreizler}, {Lopez-Morales},
  {Husser}, {Schmidt}, \& {Collet}}]{Khalafinejad+2017}
{Khalafinejad}, S., {von Essen}, C., {Hoeijmakers}, H.~J., {et~al.} 2017, \aap,
  598, A131

\bibitem[{{Koskinen} {et~al.}(2013){Koskinen}, {Harris}, {Yelle}, \&
  {Lavvas}}]{Koskinen+2013}
{Koskinen}, T.~T., {Harris}, M.~J., {Yelle}, R.~V., \& {Lavvas}, P. 2013,
  Icarus, 226, 1678

\bibitem[{{Kurucz}(1992)}]{1992RMxAA..23...45K}
{Kurucz}, R.~L. 1992, Rev. Mexicana Astron. Astrofis., 23, 45

\bibitem[{{Lammer} {et~al.}(2003){Lammer}, {Selsis}, {Ribas}, {Guinan},
  {Bauer}, \& {Weiss}}]{Lammer+2003}
{Lammer}, H., {Selsis}, F., {Ribas}, I., {et~al.} 2003, \apjl, 598, L121

\bibitem[{{Lodders}(2003)}]{2003ApJ...591.1220L}
{Lodders}, K. 2003, \apj, 591, 1220

\bibitem[{{Lothringer} {et~al.}(2020){Lothringer}, {Fu}, {Sing}, \&
  {Barman}}]{Lothringer+2020}
{Lothringer}, J.~D., {Fu}, G., {Sing}, D.~K., \& {Barman}, T.~S. 2020, \apjl,
  898, L14

\bibitem[{{Narita} {et~al.}(2005){Narita}, {Suto}, {Winn}, {Turner}, {Aoki},
  {Leigh}, {Sato}, {Tamura}, \& {Yamada}}]{Narita+2005b}
{Narita}, N., {Suto}, Y., {Winn}, J.~N., {et~al.} 2005, \pasj, 57, 471

\bibitem[{{Noguchi} {et~al.}(2002){Noguchi}, {Aoki}, {Kawanomoto}, {Ando},
  {Honda}, {Izumiura}, {Kambe}, {Okita}, {Sadakane}, {Sato}, {Tajitsu},
  {Takada-Hidai}, {Tanaka}, {Watanabe}, \& {Yoshida}}]{Noguchi+2002}
{Noguchi}, K., {Aoki}, W., {Kawanomoto}, S., {et~al.} 2002, \pasj, 54, 855

\bibitem[{{Pino} {et~al.}(2018){Pino}, {Ehrenreich}, {Wyttenbach}, {Bourrier},
  {Nascimbeni}, {Heng}, {Grimm}, {Lovis}, {Malik}, {Pepe}, \&
  {Piotto}}]{2018A&A...612A..53P}
{Pino}, L., {Ehrenreich}, D., {Wyttenbach}, A., {et~al.} 2018, \aap, 612, A53

\bibitem[{{Redfield} {et~al.}(2008){Redfield}, {Endl}, {Cochran}, \&
  {Koesterke}}]{Redfield+2008}
{Redfield}, S., {Endl}, M., {Cochran}, W.~D., \& {Koesterke}, L. 2008, \apjl,
  673, L87

\bibitem[{{Seager} \& {Sasselov}(2000)}]{Seager&Sasselov2000}
{Seager}, S., \& {Sasselov}, D.~D. 2000, \apj, 537, 916

\bibitem[{{Sedaghati} {et~al.}(2017){Sedaghati}, {Boffin}, {MacDonald},
  {Gandhi}, {Madhusudhan}, {Gibson}, {Oshagh}, {Claret}, \&
  {Rauer}}]{Sedaghati+2017}
{Sedaghati}, E., {Boffin}, H. M.~J., {MacDonald}, R.~J., {et~al.} 2017, \nat,
  549, 238

\bibitem[{{Seidel} {et~al.}(2020{\natexlab{a}}){Seidel}, {Ehrenreich}, {Pino},
  {Bourrier}, {Lavie}, {Allart}, {Wyttenbach}, \& {Lovis}}]{Seidel+2020}
{Seidel}, J.~V., {Ehrenreich}, D., {Pino}, L., {et~al.} 2020{\natexlab{a}},
  \aap, 633, A86

\bibitem[{{Seidel} {et~al.}(2019){Seidel}, {Ehrenreich}, {Wyttenbach},
  {Allart}, {Lendl}, {Pino}, {Bourrier}, {Cegla}, {Lovis}, {Barrado},
  {Bayliss}, {Astudillo-Defru}, {Deline}, {Fisher}, {Heng}, {Joseph}, {Lavie},
  {Melo}, {Pepe}, {S{\'e}gransan}, \& {Udry}}]{Seidel+2019}
{Seidel}, J.~V., {Ehrenreich}, D., {Wyttenbach}, A., {et~al.} 2019, \aap, 623,
  A166

\bibitem[{{Seidel} {et~al.}(2020{\natexlab{b}}){Seidel}, {Ehrenreich},
  {Bourrier}, {Allart}, {Attia}, {Hoeijmakers}, {Lendl}, {Linder},
  {Wyttenbach}, {Astudillo-Defru}, {Bayliss}, {Cegla}, {Heng}, {Lavie},
  {Lovis}, {Melo}, {Pepe}, {dos Santos}, {S{\'e}gransan}, \&
  {Udry}}]{Seidel+2020b}
{Seidel}, J.~V., {Ehrenreich}, D., {Bourrier}, V., {et~al.} 2020{\natexlab{b}},
  \aap, 641, L7

\bibitem[{{Seidel} {et~al.}(2021){Seidel}, {Ehrenreich}, {Allart},
  {Hoeijmakers}, {Lovis}, {Bourrier}, {Pino}, {Wyttenbach}, {Adibekyan},
  {Alibert}, {Borsa}, {Casasayas-Barris}, {Cristiani}, {Demangeon}, {Di
  Marcantonio}, {Figueira}, {Gonz{\'a}lez Hern{\'a}ndez}, {Lillo-Box},
  {Martins}, {Mehner}, {Molaro}, {Nunes}, {Palle}, {Pepe}, {Santos}, {Sousa},
  {Sozzetti}, {Tabernero}, \& {Zapatero Osorio}}]{Seidel+2021}
{Seidel}, J.~V., {Ehrenreich}, D., {Allart}, R., {et~al.} 2021, \aap, 653, A73

\bibitem[{{Sing} {et~al.}(2013){Sing}, {Lecavelier des Etangs}, {Fortney},
  {Burrows}, {Pont}, {Wakeford}, {Ballester}, {Nikolov}, {Henry}, {Aigrain},
  {Deming}, {Evans}, {Gibson}, {Huitson}, {Knutson}, {Showman}, {Vidal-Madjar},
  {Wilson}, {Williamson}, \& {Zahnle}}]{Sing+2013}
{Sing}, D.~K., {Lecavelier des Etangs}, A., {Fortney}, J.~J., {et~al.} 2013,
  \mnras, 436, 2956

\bibitem[{{Smette} {et~al.}(2015){Smette}, {Kausch}, {Sana}, {Noll}, {Horst},
  {Kimeswenger}, {Barden}, {Szyszka}, {Jones}, {Gallene}, {Vinther},
  {Ballester}, \& {Kerber}}]{Smette+2015}
{Smette}, A., {Kausch}, W., {Sana}, H., {et~al.} 2015, {Molecfit: Telluric
  absorption correction tool}

\bibitem[{{Snellen} {et~al.}(2008){Snellen}, {Albrecht}, {de Mooij}, \& {Le
  Poole}}]{Snellen+2008}
{Snellen}, I.~A.~G., {Albrecht}, S., {de Mooij}, E.~J.~W., \& {Le Poole}, R.~S.
  2008, \aap, 487, 357

\bibitem[{{Southworth} {et~al.}(2020){Southworth}, {Bohn}, {Kenworthy},
  {Ginski}, \& {Mancini}}]{Southworth+2020}
{Southworth}, J., {Bohn}, A.~J., {Kenworthy}, M.~A., {Ginski}, C., \&
  {Mancini}, L. 2020, \aap, 635, A74

\bibitem[{{Tabernero} {et~al.}(2021){Tabernero}, {Zapatero Osorio}, {Allart},
  {Borsa}, {Casasayas-Barris}, {Demangeon}, {Ehrenreich}, {Lillo-Box}, {Lovis},
  {Pall{\'e}}, {Sousa}, {Rebolo}, {Santos}, {Pepe}, {Cristiani}, {Adibekyan},
  {Allende Prieto}, {Alibert}, {Barros}, {Bouchy}, {Bourrier}, {D'Odorico},
  {Dumusque}, {Faria}, {Figueira}, {G{\'e}nova Santos}, {Gonz{\'a}lez
  Hern{\'a}ndez}, {Hojjatpanah}, {Lo Curto}, {Lavie}, {Martins}, {Martins},
  {Mehner}, {Micela}, {Molaro}, {Nunes}, {Poretti}, {Seidel}, {Sozzetti},
  {Su{\'a}rez Mascare{\~n}o}, {Udry}, {Aliverti}, {Affolter}, {Alves}, {Amate},
  {Avila}, {Bandy}, {Benz}, {Bianco}, {Broeg}, {Cabral}, {Conconi}, {Coelho},
  {Cumani}, {Deiries}, {Dekker}, {Delabre}, {Fragoso}, {Genoni}, {Genolet},
  {Hughes}, {Knudstrup}, {Kerber}, {Landoni}, {Lizon}, {Maire}, {Manescau}, {Di
  Marcantonio}, {M{\'e}gevand}, {Monteiro}, {Monteiro}, {Moschetti}, {Mueller},
  {Modigliani}, {Oggioni}, {Oliveira}, {Pariani}, {Pasquini}, {Rasilla},
  {Redaelli}, {Riva}, {Santana-Tschudi}, {Santin}, {Santos}, {Segovia},
  {Sosnowska}, {Span{\`o}}, {Tenegi}, {Iwert}, {Zanutta}, \&
  {Zerbi}}]{Tabernero+2021}
{Tabernero}, H.~M., {Zapatero Osorio}, M.~R., {Allart}, R., {et~al.} 2021,
  \aap, 646, A158

\bibitem[{{Vidal-Madjar} {et~al.}(2011){Vidal-Madjar}, {Sing}, {Lecavelier Des
  Etangs}, {Ferlet}, {D{\'e}sert}, {H{\'e}brard}, {Boisse}, {Ehrenreich}, \&
  {Moutou}}]{Vidal-Madjar+2011}
{Vidal-Madjar}, A., {Sing}, D.~K., {Lecavelier Des Etangs}, A., {et~al.} 2011,
  \aap, 527, A110

\bibitem[{{von Essen} {et~al.}(2020){von Essen}, {Mallonn}, {Hermansen},
  {Nixon}, {Madhusudhan}, {Kjeldsen}, \&
  {Tautvai{\v{s}}ien{\.{e}}}}]{vonEssen+2020}
{von Essen}, C., {Mallonn}, M., {Hermansen}, S., {et~al.} 2020, \aap, 637, A76

\bibitem[{{{\v{Z}}{\'a}k} {et~al.}(2019){{\v{Z}}{\'a}k}, {Kab{\'a}th},
  {Boffin}, {Ivanov}, \& {Skarka}}]{Zak+2019}
{{\v{Z}}{\'a}k}, J., {Kab{\'a}th}, P., {Boffin}, H. M.~J., {Ivanov}, V.~D., \&
  {Skarka}, M. 2019, \aj, 158, 120

\bibitem[{{West} {et~al.}(2016){West}, {Hellier}, {Almenara}, {Anderson},
  {Barros}, {Bouchy}, {Brown}, {Collier Cameron}, {Deleuil}, {Delrez}, {Doyle},
  {Faedi}, {Fumel}, {Gillon}, {G{\'o}mez Maqueo Chew}, {H{\'e}brard}, {Jehin},
  {Lendl}, {Maxted}, {Pepe}, {Pollacco}, {Queloz}, {S{\'e}gransan}, {Smalley},
  {Smith}, {Southworth}, {Triaud}, \& {Udry}}]{West+2016}
{West}, R.~G., {Hellier}, C., {Almenara}, J.~M., {et~al.} 2016, \aap, 585, A126

\bibitem[{{Winn} {et~al.}(2004){Winn}, {Suto}, {Turner}, {Narita}, {Frye},
  {Aoki}, {Sato}, \& {Yamada}}]{Winn+2004}
{Winn}, J.~N., {Suto}, Y., {Turner}, E.~L., {et~al.} 2004, \pasj, 56, 655

\bibitem[{{Wood} \& {Maxted}(2011)}]{Wood+2011}
{Wood}, P.~L., \& {Maxted}, P. F.~L. 2011, in The Astrophysics of Planetary
  Systems: Formation, Structure, and Dynamical Evolution, ed. A.~{Sozzetti},
  M.~G. {Lattanzi}, \& A.~P. {Boss}, Vol. 276, 491--492

\bibitem[{{Wyttenbach} {et~al.}(2015){Wyttenbach}, {Ehrenreich}, {Lovis},
  {Udry}, \& {Pepe}}]{Wyttenbach+2015}
{Wyttenbach}, A., {Ehrenreich}, D., {Lovis}, C., {Udry}, S., \& {Pepe}, F.
  2015, \aap, 577, A62

\bibitem[{{Wyttenbach} {et~al.}(2017){Wyttenbach}, {Lovis}, {Ehrenreich},
  {Bourrier}, {Pino}, {Allart}, {Astudillo-Defru}, {Cegla}, {Heng}, {Lavie},
  {Melo}, {Murgas}, {Santerne}, {S{\'e}gransan}, {Udry}, \&
  {Pepe}}]{Wyttenbach+2017}
{Wyttenbach}, A., {Lovis}, C., {Ehrenreich}, D., {et~al.} 2017, \aap, 602, A36

\bibitem[{{Wyttenbach} {et~al.}(2020){Wyttenbach}, {Molli{\`e}re},
  {Ehrenreich}, {Cegla}, {Bourrier}, {Lovis}, {Pino}, {Allart}, {Seidel},
  {Hoeijmakers}, {Nielsen}, {Lavie}, {Pepe}, {Bonfils}, \&
  {Snellen}}]{Wyttenbach+2020}
{Wyttenbach}, A., {Molli{\`e}re}, P., {Ehrenreich}, D., {et~al.} 2020, \aap,
  638, A87

\bibitem[{{Yan} {et~al.}(2017){Yan}, {Pall{\'e}}, {Fosbury}, {Petr-Gotzens}, \&
  {Henning}}]{Yan+2017}
{Yan}, F., {Pall{\'e}}, E., {Fosbury}, R.~A.~E., {Petr-Gotzens}, M.~G., \&
  {Henning}, T. 2017, \aap, 603, A73

\bibitem[{{Yelle}(2004)}]{Yelle2004}
{Yelle}, R.~V. 2004, Icarus, 170, 167

\end{thebibliography}

\end{document}